\makeatletter
\declare@file@substitution{revtex4-1.cls}{revtex4-2.cls}
\makeatother

\documentclass[twocolumn]{aastex631}

\usepackage{newtxtext,newtxmath}
\usepackage[T1]{fontenc}
\usepackage{ae,aecompl}

\usepackage{mathtools} %math tools
\usepackage{amsmath} %math package
\usepackage{bm} %bold math- preferable to \mathbf and \boldsymbol
\usepackage{amsfonts} %math fonts
\usepackage{graphicx} %enhanced support for graphics
\usepackage{bigstrut} %more spaces in tables
\usepackage{tabularx} %tables with adjustable-width columns
\usepackage{cancel} %Lines through math formulae
\usepackage{xcolor}
\usepackage{soul}
\usepackage{upgreek}
\usepackage{slashed} %command for Feynman slash notation
\usepackage{epstopdf} %converts .eps or .ps files to pdf format for figures
\newcommand\rme{{\rm e}} %Roman lowercase e
\newcommand\rmd{{\rm d}} %Roman lowercase d
\newcommand\mue{\mu_{\rm e}} %command \mue for electron chemical potential
\usepackage[subnum]{cases} %special equation numbering
  {\color{red}}%
  {}
  {\color{blue}}%
  {}

\shorttitle{Magnetar magnetothermal evolution with Landau-quantized electrons}
\shortauthors{P.~B. Rau}

\begin{document}

\title{Magnetar magnetothermal evolution with Landau-quantized electrons: enhanced Joule heating due to free electron diamagnetism and the magnetar heating problem}

\correspondingauthor{Peter B. Rau}
\email{peter.rau@columbia.edu}

\author[0000-0001-5220-9277]{Peter B. Rau}
\affiliation{Columbia Astrophysics Laboratory, Columbia University, New York, NY 10027, USA}

%\author{Yuri Levin}
%\affiliation{Columbia Astrophysics Laboratory, Columbia University, New York, NY 10027, USA}
%\affiliation{Department of Physics and Astronomy, Monash Univeristy, Clayton, VIC 3800, Australia}


\begin{abstract}
Magnetars are systematically more luminous than other neutron stars. Previous studies suggest that supporting such high luminosities requires a heat source beyond standard Ohmic dissipation-- the ``magnetar heating problem''. We previously demonstrated using periodic box simulations that de Haas--van Alphen (dHvA) oscillations of the Landau quantization-induced magnetization in a neutron star could repeatedly generate large-gradient field components that undergo rapid Ohmic dissipation. Using the \texttt{QMFM} finite volume code developed for this purpose, we perform axisymmetric 2.5D electron magnetohydrodynamics plus thermal evolution simulations of a realistic neutron star crust, including all Landau quantization effects, to quantify this enhancement to field dissipation and heating of the star.

We show that Landau quantization effects generate strong, small-scale magnetic field structures whose dissipation can heat the field-confined equatorial hot spot beyond its temperature in the absence of Landau quantization. Enhanced heating is greatest for weaker initial fields $B\sim10^{14}$ G, for which magnetic energy is dissipated 30\% faster: in this case, the crust cools sufficiently for the dHvA oscillations to reach large amplitudes. For the stronger fields $\gtrsim B\sim5\times10^{14}$ G required to power the most luminous magnetars, the enhancement is modest, as dissipation of these fields heats the crust sufficiently to thermally suppress dHvA oscillations. In cases where heating is significantly enhanced, most of the heat goes into increasing the neutrino luminosity, and surface photon luminosity is only slightly increased. For the simple crust-confined fields that we simulated, Landau quantization-enhanced Joule heating is thus insufficient to explain the magnetar heating problem.

\end{abstract}


\section{Introduction}
\label{sec:Introduction}

The evolution of the interior magnetic field of neutron stars is of long-standing astrophysical interest due to its connection with the exterior magnetospheric field, so the interior evolution will leave an imprint on the radiation emitted from within the magnetosphere. For the most strongly-magnetized neutron stars, the magnetars, interior field evolution is likely also responsible for initiating energetic X-ray/gamma ray outbursts and giant flares, with strong Lorentz forces causing crust failures or ejecting parts of the crust into the magnetosphere. Because field dissipation is strongly temperature-dependent and itself heats the crust, long-term magnetic field evolution simulations should ideally be coupled to the interior thermal evolution. For strong magnetic fields like those in magnetars, heat conduction will be strongly anisotropic and preferentially along magnetic field lines, leading to anisotropic interior and surface temperature profiles.

The magnetic field evolution is of fundamentally different nature in the superconducting fluid core and solid crust of a neutron star. The timescale of core field evolution is generally believed to be on the order $>1$ Myr ~\citep{Gusakov2019,Goglichidze2025}: for this reason, many neutron star magnetic field evolution simulations focus on the crust, which can evolve over much shorter timescales. In the crust, magnetic field evolution is governed by electron magnetohydrodynamics (electron MHD), in which the electrons move with respect to a static ionic lattice. Electron MHD in neutron star crusts is a well-studied topic~\citep{Hollerbach2002,Cumming2004,Pons2007,Pons2009,Vigano2012,Kojima2012,Vigano2013,Gourgouliatos2013,Gourgouliatos2014,
Gourgouliatos2015,Gourgouliatos2016,Bransgrove2018,Gourgouliatos2020,Vigano2021,Gourgouliatos2022,Dehman2023}, including for magnetar-strength fields $\sim10^{14}$--$10^{15}$ G. The assumptions underlying electron MHD are believed to break down for very strong magnetic fields: if the Lorentz force exceeds the maximum elastic stress the crust can support, the crust can undergo plastic failure and the nuclear lattice can no longer be assumed stationary. Plastic failure of the crust does not appear to completely suppress the Hall effect~\citep{Gourgouliatos2021}, and so a description of a magnetar crust's magnetic evolution based on electron MHD should still be valid above the crust breaking field strength.

A complete study of crustal magnetic field evolution should also consider the star's thermal evolution, since the microphysics of the magnetic field evolution (e.g., electrical conductivity) is temperature-dependent. The thermal evolution of neutron stars is reviewed in e.g.,~\citet{Potekhin2015}. The main output of thermal evolution studies is stellar surface luminosity versus time curves, to compare to existing neutron star surface X-ray luminosity observations. Thermal evolution-only simulations~\citep{Gnedin2001,Page2004,Geppert2004,Page2007,Aguilera2008,Page2009,Ho2015} allow neutron star interior physics to be constrained through their effects on the neutrino emission and heat capacity of the star. Reasonable agreement with observations has been obtained for most classes of neutron star, with the exception of the magnetars. 
 
Joint magnetothermal evolution studies in neutron star crusts~\citep{Pons2009,Vigano2013,DeGrandis2020,Vigano2021,Igoshev2021,Dehman2023,Dehman2023a,Ascenzi2024,Suvorov2026} are necessary to accurately model the thermal luminosity of magnetars, whose magnetic field significantly changes the thermal evolution: the neutrino emissivity, thermal conductivity, and Joule heating of the crust are all field-dependent. The magnetars are found to be systematically hotter than other classes of neutron star, with redshifted surface luminosities one-to-two orders of magnitude larger than those of other neutron stars between ages $\sim10^3$--$10^5$ yr.~\citet{Vigano2013} argued that the higher surface temperatures of magnetars can be explained by a combination of (1) Ohmic dissipation of strong magnetic fields aided by the Hall effect, which generates small-scale field structure and hence promotes  dissipation; (2) light element envelopes formed of accreted matter which increase the thermal conductivity in the outer layer of the star. However, light elements will be quickly burned away at hot surface of magnetars~\citep{Chang2004,Chang2010}, and their young age means that it is unlikely that they will be able to replenish these light elements through accretion. Thermal evolution-only simulations of magnetars by~\citet{Potekhin2018}, which included magnetic field effects through a fixed, uniform magnetic field magnitude but without Joule heating, found that the increased thermal conductivity in a quantizing magnetic field, and radiation from a magnetically condensed surface~\citep{Ruderman1971,Lai1997,Medin2007}, enhance the surface luminosity sufficiently to explain the luminosity of around half the magnetars, but is insufficient to explain the most luminous magnetars. 

%These simulations also accounted for neutron superfluidity in the inner crust, which by reducing the neutrino emissivity helps keep kyr-age neutron stars warmer for longer, and core neutron superfludity-proton superconductivity, which help the star remain hotter for longer by increasing the baryonic specific heat capacity and reducing the Urca neutrino emissivity~\citep{Ho2012}.

It is now apparent that some heat source is needed to power the magnetar surface luminosity. If internal heating is responsible for the higher observed surface temperatures, it must occur within the crust and predominantly in the outer crust and/or outer part of the inner crust. Otherwise, the heat will simply flow into the core and be efficiently radiated away as neutrinos~\citep{Kaminker2006,Ho2012}.~\citet{Beloborodov2016} examined four possible mechanisms to explain the ``magnetar heating problem'': ambipolar diffusion heating of the core, mechanical dissipation of the crust stressed beyond the elastic limit, Ohmic dissipation of the crustal magnetic field and surface bombardment by accelerated charged particles. They found that the second and fourth mechanisms were insufficient to explain the observed luminosities, and that a magnetar heated by ambipolar diffusion in its core would not have a lifetime consistent with observations. The ambipolar diffusion heating scenario was re-examined in~\citet{Tsuruta2023}, who found it could be consistent with observation in the case of light element contamination of the crust, though a more detailed calculation by~\citet{Moraga2025}, which simulated the magnetic field and consistently solved for the core fluid motion, disfavors the scenario as a sole explanation for the magnetar heating problem. For the crustal Ohmic dissipation scenario to be able to power the observed surface luminosities without a light element envelope,~\citet{Beloborodov2016} argued that fields changing by $\gtrsim 10^{16}$ G over length scales $\lesssim 10^2$ cm would be required. 

In the strong magnetic fields inside magnetars and in their magnetospheres, electrons will be quantized into Landau levels~\citep{Canuto1971,Harding2006}. The cyclotron orbits of the quantized electrons generates a magnetization field $\bm{M}$ to oppose the applied magnetic field in a phenomenon known as free electron diamagnetism~\citep{Darwin1931,Sondheimer1951}, and this magnetization field undergoes de Haas--van Alphen (dHvA) oscillations. In a previous paper~\citep{Rau2023}, we argued that Ohmic dissipation of the crustal magnetic field could be enhanced by these dHvA oscillations, particularly those of the differential magnetic susceptibility $\chi_{\mu}$. Unlike the oscillations of the magnetization, which are a small fraction of the magnetic field $\bm{B}$, the oscillations of the differential magnetic susceptibility (a dimensionless quantity) can be of order $1/(4\pi)$. This effect is expected to persist for a wide range of temperatures, densities and field strengths. In~\citet{Rau2025} (henceforth RW25), we used the spectral method solver \texttt{Dedalus} to study electron MHD including Landau quantization-induced magnetization in periodic box simulations meant to represent local evolution within a magnetar crust. These simulations demonstrated an enhancement of Ohmic dissipation of the magnetic field due to the continuous generation of small-scale, rapidly-dissipating magnetic field features by the dHvA oscillations of $M$ and the magnetic susceptibilities. However, the simulations in RW25 neglected thermal evolution, which provides a key feedback role by thermally suppressing Landau quantization effects, and assumed uniform electron densities, while the electron density changes by two orders of magnitude across the neutron star inner crust alone. 

In this paper, we use the finite volume code \texttt{QMFM} (Quantizing Magnetic Fields in Magnetars), developed for this project, to study the joint magnetothermal evolution in a neutron star crust. This code differs from previous ones used for studying this problem in two main aspects: (1) it is a fully MPI-parallelized finite volume code, and (2) it incorporates the effects of Landau quantization in both transport and thermodynamics, including the magnetization and its derivatives, whereas previously only its effect on transport had been considered. We focus on examining the effect of Landau quantization on the observed surface luminosity of magnetars, in particular the unexplored effect of the dHvA oscillations of the magnetic susceptibilities. The simulations are done in 2.5D axisymmetric spherical coordinates. To make the numerics tractable, we use certain simplifying approximations that leave the key novel physics intact and allow us to assess its importance to magnetothermal evolution, including assuming that electron MHD is valid throughout the simulation. The primary goal of this work is to compare simulations with and without Landau quantization effects included to determine their importance to magnetar magnetothermal evolution.

In Section~\ref{sec:SimEquations}, we briefly outline the equations we solved in our simulations, the input crust model and the microphysics. Section~\ref{sec:BCICApprox} discusses the choices of initial and boundary conditions we use in the simulations, and the set of approximations used in computing the various thermodynamic functions for Landau-quantized electrons. Section~\ref{sec:QMFM} describes the \texttt{QMFM} code, and in Section~\ref{sec:SimResults} we discuss our simulations including novel quantization effects. The implications of these results are discussed in Section~\ref{sec:Conclusion}. We work in Gaussian units throughout.

\section{Magnetothermal evolution equations in strongly-magnetized neutron star crusts}
\label{sec:SimEquations}

We consider the joint magnetothermal evolution of a neutron star crust in a fixed general relativistic spacetime, using the formalism for inclusion of spacetime curvature terms using an orthonormal basis as described in~\citet{Radler2001}. The background spacetime metric is of Schwarzschild form
\begin{equation}
\rmd s^2 = -\rme^{\nu}\rmd t^2 + \rme^{\lambda}\rmd r^2 + r^2(\rmd\theta^2+\sin^2\theta\rmd\phi^2),
\end{equation} 
where $\nu$ and $\lambda$ are functions of radial coordinate $r$ only: $\rme^{\lambda}=(1-2GM(r)/(c^2r))^{-1}$ for enclosed mass $M(r)$, while $\nu$ depends on the details of the crust model and is solved alongside the Tolman--Oppenheimer--Volkoff equation. We assume the magnetic field is confined to the crust, a common approximation (e.g.,~\citet{Hollerbach2002,Rheinhardt2004,Pons2007,Gourgouliatos2014,Dehman2023}), which allows us to limit our magnetic field simulation to only this region. This in effect is assuming that the neutron star core is type-I superconducting and hence expels the magnetic field. This is not realistic, since the first critical field $H_{c1}$ for type-I superconducting protons at the crust-core boundary is of order $10^{15}$ G~\citep{Sedrakian2019}. $H_{c1}$ could thus easily be exceeded by magnetar-strength fields, but even if it is not, the timescale for expulsion of fields below this may be of order the age of the universe~\citep{Baym1969a}. Regardless, as a first approximation we maintain this assumption. This also allows us to limit our spatial resolution to the crust only, which helps us better resolve the dHvA oscillations whose effects on the magnetothermal evolution are the main focus of this paper. We additionally assume that the exterior of the star is a current-free vacuum, and thus that the magnetic field here is a potential field.

In a neutron star crust the electric current is due to the electrons moving relative to the approximately stationary, neutralizing nuclear lattice. Under these conditions, the Hall term appears in Ohm's Law, and the corresponding magnetohydrodynamics are called electron MHD. Accounting for anisotropic conductivity, but ignoring the electron inertia, the electron pressure gradient and thermoelectric terms, Ohm's Law thus takes the form
\begin{equation}
\bm{E}'=\bm{E}+\frac{1}{c}\bm{v}\times\bm{B}=\frac{1}{n_{\text{e}}ec}\bm{J}\times\bm{B}+\rho_{\parallel}\bm{J}_{\parallel}+\rho_{\perp}\bm{J}_{\perp}.
\label{eq:OhmsLaw}
\end{equation}
$\bm{E}'$ is the electric field in the rest frame of the crustal lattice which moves with velocity $\bm{v}$, $\bm{E}$ is the electric field in the frame where the crustal lattice moves with velocity $\bm{v}$, and $\bm{J}$ is the (free) electric current density. $n_{\rme}$ is the electron number density and $e>0$ the charge of the proton. $\rho_{\parallel}$ and $\rho_{\perp}$ are the electrical resistivities parallel to and perpendicular to the magnetic field; the components of $\bm{J}$ projected parallel to and perpendicular to the magnetic field are
\begin{align}
\bm{J}_{\parallel}=(\hat{\bm{B}}\cdot\bm{J})\hat{\bm{B}}, && \bm{J}_{\perp}=(\bm{I}-\hat{\bm{B}}\hat{\bm{B}})\cdot\bm{J},
\end{align}
where $\hat{\bm{B}}=\bm{B}/B$ is the unit vector in the direction of $\bm{B}$ and $\bm{I}$ is the identity tensor. $\rho_{\parallel}\neq\rho_{\perp}$ is a result of Landau quantization of the electrons, which introduces distinct relaxation times perpendicular $\tau_{\perp}$ and parallel $\tau_{\parallel}$ to the magnetic field~\citep{Potekhin1999a}. In the non-quantizing case $\tau=\tau_{\parallel}=\tau_{\perp}$ and we would have $\rho_{\parallel}=\rho_{\parallel}=\sigma_{\parallel}^{-1}$, but for the conductivities we have $\sigma_{\perp}=\sigma_{\parallel}/(1+(\omega_g\tau)^2)$ where $\omega_g$ is the electron gyrofrequency. In magnetars, $(\omega_g\tau)\gg 1$ and so $\sigma_{\perp}\ll\sigma_{\parallel}$, but this large difference in magnitude does not extend to the resistivities computed when the conductivity tensor is inverted. Unlike $\sigma_{\parallel}\gg\sigma_{\perp}$, since $\tau_{\perp}$ and $\tau_{\parallel}$ are of the same order of magnitude, so are $\rho_{\parallel}$ and $\rho_{\perp}$, though $\rho_{\parallel}<\rho_{\perp}$ is still true.

The (free) current density follows from Amp\`{e}re's Law
\begin{equation}
\bm{J}=e(Zn_{\text{N}}\bm{v}-n_{\text{e}}\bm{v}_{\text{e}})=\frac{c}{4\pi}\rme^{-\nu/2}\bm{\nabla}\times\left(\rme^{\nu/2}\bm{H}\right),
\label{eq:AmperesLaw}
\end{equation}
where $Z$ the atomic number of the nuclei and $n_{\text{N}}$ the number density of nuclei. We assume local charge neutrality $Zn_{\text{N}}=n_{\text{e}}$. $\bm{H}=\bm{B}-4\pi\bm{M}$ where $\bm{M}$ is the magnetization field. $\bm{H}$ is computed from the grand potential density $\Omega$, consisting of a non-magnetic contribution from the baryons in the ionic lattice $\Omega_{\rm b}$, from the magnetized electrons $\Omega_{\text{e}}(\mu_{\rm e},B,T)$, and from the vacuum magnetic field $B^2/(8\pi)$:
\begin{equation}
\bm{H}=4\pi\left.\frac{\partial \Omega}{\partial \bm{B}}\right|_{T,n_{\text{e}}}, \qquad
\bm{M}=-\left.\frac{\partial \Omega_{\text{e}}}{\partial B}\right|_{T,n_{\text{e}}}\bm{\hat{B}}.
\label{eq:HandM}
\end{equation}

We will only consider the case of electron MHD, where the nuclear lattice is fixed $\bm{v}=0$: otherwise we would be considering the more general Hall MHD. The magnetic induction equation governing the evolution of $\bm{B}$ is
\begin{align}
\frac{\partial\bm{B}}{\partial t}={}&-c\bm{\nabla}\times\left(\rme^{\nu/2}\bm{E}\right)
\nonumber
\\
={}&\bm{\nabla}\times\left[\frac{\rme^{\nu/2}}{en_{\text{e}}}\bm{B}\times\bm{J}\right]-c\bm{\nabla}\times\left[\rme^{\nu/2}\left(\rho_{\parallel}\bm{J}_{\parallel}+\rho_{\perp}\bm{J}_{\perp}\right)\right].
\label{eq:MagneticInduction}
\end{align}
Using Eq.~(\ref{eq:HandM}) we can write Eq.~(\ref{eq:AmperesLaw}) as
\begin{align}
\bm{J}={}&\frac{c}{4\pi}\rme^{-\nu/2}\left(1-4\pi\frac{M}{B}\right)\bm{\nabla}\times\left(\rme^{\nu/2}\bm{B}\right)
\nonumber
\\
{}&-c\left(\bm{\nabla}M-\frac{M}{B}\bm{\nabla}B\right)\times\bm{\hat{B}}.
\label{eq:JExpansion1}
\end{align}
To make explicit the contributions from the differential magnetic susceptibility $\chi_{\mu}$ and the two mixed susceptibilities $\mathcal{M}_{\mu}$ and $\mathcal{M}_T$, defined as
\begin{align}
\chi_{\mu}\equiv\left.\frac{\partial M}{\partial B}\right|_{T,\mue}, && \mathcal{M}_{\mu}\equiv\left.\frac{\partial M}{\partial \mue}\right|_{T,B}, &&
\mathcal{M}_T\equiv\left.\frac{\partial M}{\partial T}\right|_{\mue,B}, 
\label{eq:MagnetizationPartialDerivatives}
\end{align}
we write
\begin{equation}
\bm{\nabla}M = \chi_{\mu}\bm{\nabla}B+\mathcal{M}_{\mu}\bm{\nabla}\mue+\mathcal{M}_T\bm{\nabla}T.
\end{equation}
Using this Eq.~(\ref{eq:JExpansion1}) can be simplified to
\begin{align}
\bm{J}={}&\frac{c}{4\pi}\rme^{-\nu/2}\left(1-4\pi\frac{M}{B}\right)\bm{\nabla}\times(\rme^{\nu/2}\bm{B})
\nonumber
\\
{}&-c\left(\left(\chi_{\mu}-\frac{M}{B}\right)\bm{\nabla}B+\mathcal{M}_{\mu}\bm{\nabla}\mue+\mathcal{M}_T\bm{\nabla}T\right)\times\bm{\hat{B}}.
\label{eq:JMagnetization}
\end{align}
This procedure to explicitly include $\chi_{\mu}$, $\mathcal{M}_{\mu}$ and $\mathcal{M}_T$ in the evolution equations must be used if we want to study the effect of the large amplitude dHvA oscillations of $\chi_{\mu}$ without using extremely fine spatial grid spacing in our numerical implementation. Otherwise, the finite differencing of $M$ would not capture the large oscillations of the derivatives of $M$. This form also demonstrates that deviations from the $\bm{B}=\bm{H}$ result only contribute to the current density perpendicular to $\bm{B}$.

To determine the temperature evolution, we use the generalized heat equation without entropy advection
\begin{equation}
\tilde{T}\frac{\partial s}{\partial t}=-\bm{\nabla}\cdot(\rme^{\nu}\bm{q})+\rme^{\nu}(\dot{h}-\dot{q}),
\label{eq:GeneralizedHeatEquationStart}
\end{equation}
where $s$ is the volumetric entropy density, $\tilde{T}=\rme^{\nu/2}T$ is the gravitationally-redshifted temperature in terms of the local temperature $T$, $\bm{q}$ is the heat flux density, and $\dot{h}$ and $-\dot{q}$ are the volumetric heating and cooling rates (we take $\dot{q}>0$). The heat source is Joule heating and the cooling is provided by neutrino emission:
\begin{align}
\dot{h}={}&\bm{J}\cdot\bm{E}=\rho_{\perp}J^2_{\perp}+\rho_{\parallel}J^2_{\parallel},
\\
\dot{q}={}&\dot{q}_{\nu}.
\end{align}
Eq.~(\ref{eq:GeneralizedHeatEquationStart}) is then transformed into an evolution equation for the temperature; as our system includes $B$-dependence in the entropy density, we show this calculation explicitly. The entropy density can be split into contributions from the electrons and from the baryons (nuclear lattice plus dripped superfluid neutrons in the inner crust). We write
\begin{equation}
s=s_{\rme}(T,\mue,B)+s_{\rm b}(T),
\end{equation}
where $s_{\rme}$ and $s_{\rm b}$ are the electron and baryon entropy density respectively. Since we assume the lattice and dripped neutrons are stationary and not affected by the magnetic field, we can ignore non-temperature dependence in $s_{\rm b}$. The differential of $s$ is
\begin{align}
\text{d}s={}&\left.\frac{\partial s_{\text{e}}}{\partial \mu_{\rm e}}\right|_{T,B}\rmd \mu_{\rm e}+\left.\frac{\partial s_{\rm e}}{\partial T}\right|_{\mu_{\rm e},B}\rmd T+\left.\frac{\partial s_{\rm e}}{\partial B}\right|_{\mu_{\rm e},T}\rmd B+\frac{\partial s_{\rm b}}{\partial T}\rmd T,
\nonumber
\\
={}&\left.\frac{\partial s_{\rm e}}{\partial \mu_{\rme}}\right|_{T,B}\rmd \mu_{\rme}+\frac{c_V}{T}\rmd T+\mathcal{M}_T\rmd B,
\end{align}
where $c_V=c_{V,{\rm e}}+c_{V,{\rm b}}$ is the total volumetric specific heat capacity of the electrons and baryons at fixed magnetic field. A Maxwell relation gives the equivalence of the final equation in Eq.~(\ref{eq:MagnetizationPartialDerivatives}) and $(\partial s_{\text{e}}/\partial B)_{\mue,T}$; $\mathcal{M}_T$ is interpreted as the magnetocaloric coefficient. The time derivative of $s$ is hence
\begin{equation}
\frac{\partial s}{\partial t}=\left.\frac{\partial s_{\rme}}{\partial \mue}\right|_{T,B}\frac{\partial \mue}{\partial t}+\frac{c_V}{T}\frac{\partial T}{\partial t}+\mathcal{M}_T\hat{\bm{B}}\cdot\frac{\partial \bm{B}}{\partial t},
\label{eq:dsdt}
\end{equation}
where we will hold the electron chemical potential constant during the evolution, so the first term on the right-hand side of Eq.~(\ref{eq:dsdt}) vanishes. The generalized heat equation thus becomes, also writing $\dot{h}$ and $\dot{q}$ explicitly,
\begin{align}
c_V\frac{\partial \tilde{T}}{\partial t} + \tilde{T}\mathcal{M}_T\hat{\bm{B}}\cdot\frac{\partial 
\bm{B}}{\partial t}={}& -\bm{\nabla}\cdot(\rme^{\nu}\bm{q})
\nonumber
\\
{}&+\rme^{\nu}\left(\rho_{\perp}J^2_{\perp}+\rho_{\parallel}J^2_{\parallel}-\dot{q}_{\nu}\right),
\label{eq:GeneralizedHeatEquation}
\end{align}
After substitution of $\partial \bm{B}/\partial t$ with Eq.~(\ref{eq:MagneticInduction}) and then moving this term to the right-hand side of the equation, Eq.~(\ref{eq:GeneralizedHeatEquation}) is reduced to an evolution equation for $\tilde{T}$ only. 

Since the thermal conductivity is dominated by the electrons, the thermal conductivity tensor $\bm{\kappa}$ can be strongly anisotropic in strong magnetic fields, with heat flow along the field lines strongly enhanced compared to perpendicular to field lines. The heat flux density $\bm{q}$ is thus written in terms of $\bm{\kappa}$ and the gradient of $\tilde{T}$ as~\citep{Geppert2004,Perez-Azorin2006}
\begin{align}
\bm{q}={}&-\rme^{-\nu/2}\bm{\kappa}\cdot\bm{\nabla}\tilde{T}
\nonumber
\\
={}&-\rme^{-\nu/2}\kappa_{\parallel}\bm{\hat{B}}(\bm{\hat{B}}\cdot\bm{\nabla}\tilde{T})-\rme^{-\nu/2}\kappa_{\perp}\left(\bm{I}-\bm{\hat{B}}\bm{\hat{B}}\right)\cdot\bm{\nabla}\tilde{T}
\nonumber
\\
{}&-\rme^{-\nu/2}\kappa_{\wedge}\bm{\hat{B}}\times\bm{\nabla}\tilde{T},
\end{align}
where $\kappa_{\parallel}$ and $\kappa_{\perp}$ are the thermal conductivities parallel and perpendicular to the magnetic field and $\kappa_{\wedge}$ is the Hall thermal conductivity, equivalent to the Leduc--Righi coefficient divided by $B$. These conductivities are related to each other by
\begin{subequations}
\begin{align}
\kappa_{\perp} = \frac{\tau_{\perp}}{\tau_{\parallel}}\frac{\kappa_{\parallel}}{1+(\tau_{\perp}\omega_g)^2}, \qquad
\kappa_{\wedge} = (\tau_{\perp}\omega_g)\kappa_{\perp}
\end{align}
\end{subequations}
Since $\tau_{\perp}$ and $\tau_{\parallel}$ are of the same order of magnitude~\citep{Potekhin1999a}, heat conduction is predominantly along the magnetic field lines as long as $(\omega_g\tau_{\perp})\gg 1$.

We retain the magnetocaloric term in Eq.~(\ref{eq:GeneralizedHeatEquation}), although it is generally significantly smaller than the Joule heating term. The ratio of these two terms is approximately
\begin{align}
{}&\frac{\rme^{\nu}\rho J^2}{\tilde{T}\mathcal{M}_T\hat{\bm{B}}\cdot\frac{\partial\bm{B}}{\partial t}}\approx \frac{en_{\rme}c}{4\pi\sigma\mathcal{M}_TT}
\nonumber
\\
\approx{}& 30\left(\frac{100\ {\rm G}\ {\rm K}^{-1}}{\mathcal{M}_T}\right)\left(\frac{4\times 10^8\ {\rm K}}{T}\right)\left(\frac{n_{\rme}}{10^{35}\ {\rm cm}^{-3}}\right)\left(\frac{10^{23}\ {\rm s}^{-1}}{\sigma}\right),
\end{align}
ignoring the effects of Landau quantization on the current density. $\mathcal{M}_T$, like $M$, $\chi_{\mu}$ and $\mathcal{M}_{\mu}$, undergoes dHvA oscillations; $100$ G/K is a reasonable approximation for the amplitude of these oscillations, while the spatial mean of $\mathcal{M}_T$ is roughly an order of magnitude smaller than this.

Eq.~(\ref{eq:MagneticInduction}) and~(\ref{eq:GeneralizedHeatEquation}) are the two principal equations that we evolve in our simulations. They are both nonlinear, explicitly through the Hall term in the case of the magnetic induction equation, and more generally through the nonlinear dependence on $B$ and $T$ of their coefficients. For the sake of simplicity, we do not include the thermoelectric terms: simulation of electron MHD with these included~\citep{DeGrandis2020,Gakis2024} have generally shown them to be mostly unimportant except in extreme cases where temperatures $\gtrsim 10^9$ K can be maintained in the crust over long periods. 

As in RW25, we monitor magnetic energy conservation using a modified Poynting theorem
\begin{equation}
\frac{\partial}{\partial t}\left(\rme^{\nu/2}u_B\right)=-\rme^{\nu}\bm{J}_B\cdot\bm{E}-\bm{\nabla}\cdot\left(\rme^{\nu}\bm{S}_B\right),
\label{eq:LocalEnergyConservation}
\end{equation}
where $u_B=B^2/(8\pi)$, $\bm{J}_B=c\rme^{-\nu/2}\bm{\nabla}\times(\rme^{\nu/2}\bm{B})/(4\pi)$ and $\bm{S}_B=c\bm{E}\times\bm{B}/(4\pi)$ are the magnetic energy density, current density and Poynting vector respectively when $\bm{B}=\bm{H}$.

\subsubsection{Crust model}

The electron number density $n_{\rme}$ is featured directly in the magnetic induction equation, and the various transport coefficients are functions of the density and nuclear properties (mass number $A$, atomic number $Z$, properties of the superfluid neutrons in the inner crust) that vary throughout the crust. We use a realistic crust model generated using the BSk24~\citep{Pearson2018} equation of state (EOS) to provide these inputs. The crust is taken from the $1.4M_{\odot}$ solar mass star found by solving the TOV equation with this EOS. The inner and outer radii of this crust are $11.54$ km and $12.59$ km, although we only consider the crust down to a density of $\rho=1.5\times10^{10}$ g/cm$^3$, which occurs at radius $12.40$ km. This EOS does not account for the modifications made to the crust composition by the strong magnetic field, such as different nuclear layers~\citep{Mutafchieva2019} and a different location for the neutron drip line~\citep{Chamel2015}. We do not expect that including these overall small compositional changes will significantly modify our results.

The electron number density $n_{\rme}$, set by the equation of state, is featured directly in the magnetic induction equation in the Hall diffusivity, and the various transport coefficients are functions of $n_{\rme}$. Through their dependence on $\mue$, the thermodynamic functions determining the magnetization and magnetic susceptibilities are also functions of $n_{\rme}$. $n_{\rme}$ is related to the electron contribution to the grand potential density $\Omega_{\rme}$ through
\begin{equation}
n_{\rm e}=-\left.\frac{\partial\Omega_{\rme}}{\partial\mue}\right|_{B,T}.
\label{eq:ElectronDensity}
\end{equation}
The electron chemical potential $\mue$, which is an input to compute thermodynamic functions $M$, $\chi_{\mu}$ and $\mathcal{M}_{\mu}$, is computed for the initial $B$ and $n_{\rm e}$ configuration by solving Eq.~(\ref{eq:ElectronDensity}) for $\mue$, a nonlinear equation. We then hold $\mue$ fixed throughout the simulation. In principle, $n_{\rme}$ should be updated as the simulation progresses due to the change in $B$, but outside of the strongly quantizing limit $n_{\rm max}\lesssim 2$, the small changes in $n_{\rme}$ during the evolution will negligibly affect the results.

\subsubsection{Microphysics}

As discussed in the opening paragraph, strong magnetic fields and the resulting Landau quantization of electrons have long been known to modify the thermodynamic and transport properties of matter. One goal of this work is to include as many of these modifications as possible in a full magnetothermal simulation. The effect of Landau quantization to the transport coefficients, the de Haas--van Alphen oscillations, are well-studied. For the electrical and thermal conductivity tensors, we use the results of~\citet{Potekhin1999a} with the Coulomb logarithm fitting function of~\citet{Gnedin2001}, noting at the temperatures relevant to our simulations the transport coefficients are dominated by electron-ion collisions~\citep{Potekhin2018}. We use the Wiedemann--Franz law to relate the electrical and thermal conductivity tensors. The fitting formuale of \citet{Potekhin1999a} have problematic divergences which scale as $(\nu-n_{\rm max})^{-1/2}$, where $\nu=p_F^2/(2eB)$ for electron Fermi momentum $p_F$ and magnetic field magnitude $B$, and $n_{\rm max}=\lfloor\nu\rfloor$ is the maximum occupied Landau level. Physically, these divergences would be regularized by finite temperatures, but we simply regularize them by replacing $\nu-n_{\rm max}$ with a smoothed function of $\nu$ that never equals zero, which makes the formulae suitable for use in our simulations.

In additio to electron-ion scattering, we include the contribution to thermal conductivity from electron-electron scattering~\citep{Shternin2006,Cassisi2007} and from ion-ion (phonon) scattering~\citep{Chugunov2007}. At high densities $\rho\gtrsim 10^{13}$ g/cm$^3$ and low temperatures $T\lesssim$ a few $\times10^8$ K, electron-impurity scattering dominates the electrical conductivity. We also include the electron-impurity scattering contribution to $\bm{\sigma}$ and $\bm{\kappa}$, using the impurity parameter $Q_{\rm imp}$ profile computed at the crystallization temperature for the BSk24 EoS crust from~\citet{Carreau2020a}. We also consider a pure crust $Q_{\rm imp}=0$ for comparison. We have confirmed the accuracy of our calculation of these coefficients by comparison to the publicly-available codes of A. Potekhin~\citep{Potekhin2015}.

The magnetization $M$ and its derivatives $\chi_{\mu}$ and $\mathcal{M}_{\mu}$, including the de Haas--van Alphen oscillations, are computed according to the appendix of RW25. This appendix describes the two regimes of approximation, high temperature $T>eB/(2\pi^2\mue)$ and low temperature $T<eB/(2\pi^2\mue)$, used in the numerical calculations to avoid performing costly Fermi--Dirac integrals at each time step. Temperature is important in determining whether the enhanced Ohmic dissipation effect due to dHvA oscillations is active or not, as high temperatures completely suppress the oscillations by allowing non-degenerate electrons to occupy Landau levels above the maximum value in the zero-temperature limit. 

The specific heat capacity $c_V$ consists of three parts: the Landau-quantized electrons, the nuclear lattice or ionic contribution, and in the inner crust, the superfluid neutrons. The magnetized electron contribution $c_{V,\rme}$ is computed from the grand potential density $\Omega_{\rme}$ of the electrons as
\begin{equation}
c_{V,\rme}=-T\left.\frac{\partial^2f_{\text{e}}}{\partial T^2}\right|_{B,n_{{\rme}}}.
\end{equation}
The details of this calculation, and others involving partial derivatives of $\Omega_{\rme}$, are discussed in detail in Appendix~\ref{app:Thermodynamics} and the appendix of RM25: additional partial derivatives are required in this paper compared to RM25, since we now consider evolving temperature. The ionic lattice heat capacity $c_{V,{\rm i}}$ is a combination of harmonic~\citep{Baiko2001} and anharmonic~\citep{Baiko2022} contributions. The harmonic contribution to $c_{V,{\rm i}}$ fully accounting for the effect of the magnetic field is available~\citep{Potekhin2013}, but for the crust density range included in our simulations, we have found that the magnetic effects are small and we hence use only the numerically simpler non-magnetic version. The superfluid neutron heat capacity $c_{V,{\rm nf}}$ of~\citet{Pastore2015} as described in~\citet{Potekhin2015} is used, with the neutron superfluidity control function of~\citet{Yakovlev1999} and the SFB model~\citep{Schwenk2003,Ho2015} of the $^1$S$_0$ neutron pairing gap. 

For neutrino emissivity in the crust, we include plasmon decay, neutrino brehmsstrahlung and neutrino synchrotron radiation mechanisms. The plasmon decay emissivity is taken from~\citet{Kantor2007}. The electron-nucleus neutrino brehmsstrahlung emissivity is from~\citet{Ofengeim2014}, and the neutrino synchrotron radiation emissivity is from~\citet{Bezchastnov1997}. At the densities and magnetic field strengths relevant to the simulation $10^{10}\lesssim\rho\lesssim10^{14}$ g/cm$^3$, plasmon decay is the dominant neutrino emission mechanism for $T\gtrsim 10^9$ K, neutrino synchrotron radiation dominates at $T\lesssim 10^8$ K, and neutrino brehmsstrahlung is dominant at intermediate temperatures. Modifications to neutrino emissivity due to Landau quantization of electrons mentioned in Section~\ref{sec:Introduction} are not included in this work.

\section{Boundary conditions, initial conditions and approximations}
\label{sec:BCICApprox}

\subsection{Magnetic field}
\label{sec:InitialMagneticField}

For an assumed type-I superconducting core that expels the magnetic field, we require that the radial component of the magnetic field vanishes at the inner crust-core boundary $r=R_i$. A perfectly conducting core will also have uniform electric potential, and hence the tangential components of the electric field must also vanish at this boundary. In summary, at the inner boundary we have 
\begin{subequations}
\begin{align}
B_r(r=R_i) {}&= 0,
\\
E_{\theta}(r=R_i) {}&= 0,
\\
E_{\phi}(r=R_i) {}&= 0.
\end{align}
\label{eq:InnerBC}
\end{subequations}
These inner boundary conditions guarantee that the radial Poynting vector evaluated at this surface is zero and hence no electromagnetic energy flows into/out of the core. 

The outer boundary condition at $r=R_o$ is determined by the choice of an exterior vacuum $\bm{J}=0$, which imposes that the exterior field is $\bm{B}=-\bm{\nabla}\chi$ for some scalar potential $\chi$. This requires that the radial component of $\bm{J}$ vanishes at $r=R_i$, which in axisymmetry is satisfied by imposing that the toroidal field vanishes at $r=R_i$,
\begin{equation}
B_{\phi}(r=R_o)=0.
\label{eq:OuterBCToroidal}
\end{equation}
The potential boundary condition is imposed on the poloidal components of the field constraining that $B_r$ and $B_{\theta}$ satisfy a particular relation, so $B_r$ is taken continuous at the boundary and $B_{\theta}$ is set by a spherical harmonic decomposition of $B_r$~\citep{Bransgrove2018,Dehman2023}. For a relativistic background,  this boundary condition was discussed in~\citet{Dehman2023}; we repeat it here to include the metric factors, which are an overall minor correction.

In the exterior, $\bm{B}=-\bm{\nabla}\chi$ and $\bm{\nabla}\cdot\bm{B}=0$ require that the scalar potential $\chi$ is a solution of the Laplace equation. In a Schwarzschild coordinate spacetime, which reduces to the Schwarzschild spacetime exactly in the exterior, the admissible solution which vanishes as $r\rightarrow\infty$ takes the form
\begin{equation}
\chi=-\sum_{\ell=1}a_{\ell}r^{-(\ell+1)}\left(1+\rme^{-\lambda/2}\right)^{-(2\ell+1)}Y_{\ell,0}(\theta),
\label{eq:PotentialFieldScalar}
\end{equation}
where $Y_{\ell,0}(\theta)$ are the $m=0$ spherical harmonics. 
%\begin{align}
%\chi=-\sum_{\ell=1}\bigg[{}&a_{\ell}r^{-(\ell+1)}\left(1+\rme^{-\lambda/2}\right)^{-(2\ell+1)}
%\nonumber
%\\
%{}&+b_{\ell}r^{\ell}\left(1+\rme^{-\lambda/2}\right)^{2\ell+1}\bigg]Y_{\ell,0}(\theta),
%\label{eq:PotentialFieldScalar}
%\end{align}
%The sum over spherical harmonic order $\ell$ starts at $1$ since the $\ell=0$ term must be zero from the zero divergence condition on $\bm{B}$. In the nonrelativistic limit, this reduces to a linear combination of $r^{\ell}$ and $r^{-(\ell+1)}$, the usual radial part of the solution to the Laplace equation. The requirement for $\chi$ to vanish at $r\rightarrow\infty$ implies $b_{\ell}=0$. 

In both interior and exterior of the star, we can construct $\bm{B}$ using the axisymmetric $m=0$ vector spherical harmonics
\begin{align}
\bm{B}=\sum_{\ell=1}\left[B^r_{\ell 0}(r)\bm{Y}_{\ell 0}(\theta)+B^{(1)}_{\ell 0}(r)\bm{\Psi}_{\ell 0}(\theta)+B^{(2)}_{\ell 0}(r)\bm{\Phi}_{\ell 0}(\theta)\right],
\label{eq:BVectorSpherical}
\end{align}
where
\begin{align*}
\bm{Y}_{\ell 0}=Y_{\ell 0}(\theta)\hat{\bm{r}}, 
\quad
\bm{\Psi}_{\ell 0}=\frac{\partial Y_{\ell 0}}{\partial\theta}\hat{\bm{\theta}},
\quad
\bm{\Phi}_{\ell 0}=\frac{\partial Y_{\ell 0}}{\partial\theta}\hat{\bm{\phi}}.
\end{align*}
For Eq.~(\ref{eq:BVectorSpherical}) to be divergence-free, $B^r_{\ell}(r)$ and $B^{(1)}_{\ell}(r)$ must be related by
\begin{equation}
B^{(1)}_{\ell 0}(r) = -\frac{\rme^{-\lambda(r)/2}}{\ell(\ell+1)}\frac{1}{r}\frac{\rmd}{\rmd r}\left(r^2 B^{r}_{\ell 0}(r)\right).
\label{eq:B1-Br}
\end{equation}
At $r=R_o$, we match to the potential field in the exterior, which from Eq.~(\ref{eq:PotentialFieldScalar}) has components
\begin{align}
B_r={}&\rme^{-\lambda/2}\frac{\partial\chi}{\partial r}
=\rme^{-\lambda/2}\sum_{\ell=1}a_{\ell}\frac{(\ell+1)\left(1+\rme^{-\lambda/2}\right)^{-(2\ell+1)}}{r^{\ell+2}}\nonumber
\\
{}&\qquad\qquad\times\left(1-\frac{2\ell+1}{2(\ell+1)}(\rme^{\lambda/2}-1)\right)Y_{\ell 0}(\theta),
\\
B_{\theta}={}&\frac{1}{r}\frac{\partial\chi}{\partial \theta}=-\sum_{\ell=1}\frac{a_{\ell}}{r^{\ell+2}}\left(1+\rme^{-\lambda/2}\right)^{-(2\ell+1)}\frac{\partial Y_{\ell 0}}{\partial\theta}.
\end{align}
Setting these components equal to $B_r$ and $B_{\theta}$ as given by Eq.~(\ref{eq:PotentialFieldScalar}) and using the orthonormality relations for the spherical harmonics gives
\begin{equation}
B^{(1)}_{\ell 0}(r)=-\frac{\rme^{\lambda/2}}{\ell+1}\left(1-\frac{2\ell+1}{2(\ell+1)}(\rme^{\lambda/2}-1)\right)^{-1}B^r_{\ell 0}(r),
\label{eq:B1-BrExterior}
\end{equation}
in the exterior. At $r=R_o$, we can combine this relation with Eq.~(\ref{eq:B1-Br}) to give
\begin{equation}
\left.\frac{\rmd B^r_{\ell 0}}{\rmd r}\right|_{r=R_o}=-\frac{\ell+2}{R_o}g_{\ell}\left(\rme^{\lambda(R_o)/2}\right)B^r_{\ell 0}(R_o).
\label{eq:BrellCondition}
\end{equation}
The factor
\begin{equation}
g_{\ell}\left(x\right)=\frac{2}{\ell+2}\left[\frac{\ell(\ell+1)x}{2(\ell+1)-(2\ell+1)(x-1)}+1\right],
\label{eq:g_ell}
\end{equation}
equals one in the nonrelativistic limit $x=1$. $B_r$ must satisfy Eq.~(\ref{eq:BrellCondition}), and once this is imposed, then $B_{\theta}(r=R_o)$ can be set using Eq.~(\ref{eq:B1-BrExterior}). 

For the initial field, we choose either a purely poloidal or mixed poloidal-toroidal field of the form Eq.~(\ref{eq:BVectorSpherical}), specified by choosing $B^r_{\ell 0}(r)$ and $B^{(2)}_{\ell 0}(r)$ satisfying the boundary conditions Eq.~(\ref{eq:InnerBC},\ref{eq:OuterBCToroidal},\ref{eq:BrellCondition}). $B^{(1)}_{\ell 0}(r)$ is then derived from $B^{r}_{\ell}(r)$ using Eq.~(\ref{eq:B1-Br}). 

For the radial field configuration, we consider an axisymmetric $\ell=1$ dipole field which has the form
\begin{equation}
B_r = \sqrt{\frac{4\pi}{3}}B_p\frac{R_o^2f(r)}{r^2f(R_o)}Y_{10},
\label{eq:RadialB}
\end{equation}
where $B_p$ is a constant. The function $f(r)$ is chosen to satisfy both Eq.~(\ref{eq:InnerBC}) and~(\ref{eq:BrellCondition}): we use one such form
\begin{align}
f(r) ={}& \mu r\left(j_1(\mu r) + \mathcal{B}n_1(\mu r)\right),
\\
\mathcal{B} ={}& \frac{\tan(\mu R_o)-\mathcal{F}_1}{\mathcal{F}_1\tan(\mu R_o)+1},
\\
\mathcal{F}_1 ={}& \frac{3\mu R_o(1-g_1(\rme^{\lambda(R_o)/2}))}{3(1-g_1(\rme^{\lambda(R_o)/2}))-(\mu R_o)^2},
\end{align}
which is the relativistically-corrected form of the initial radial field used initially by~\citet{Aguilera2008}. This is constructed from a sum of spherical Bessel functions $j_1(x)=\sin x/x^2-\cos x/x$ and $n_1(x)=-\cos x/x^2-\sin x/x$. $\mathcal{B}$ simplifies to $\tan(\mu R_o)$ in the nonrelativistic limit. The parameter $\mu$ is chosen to satisfy the inner boundary condition, and in the nonrelativistic limit depends only on the thickness of the crust through the ratio $R_i/R_o$. There are multiple solutions for $\mu$, with an increasing number of internal radial oscillations of $f(r)$ as $\mu$ increases. We only consider the smallest value of $\mu$ for the particular crust model used.

From Eq.~(\ref{eq:B1-Br}), the polar component of $\bm{B}$ must be
\begin{equation}
B_{\theta}=\sqrt{\frac{4\pi}{3}}B_p\rme^{-\lambda/2}\frac{R_o^2}{rf(R_o)}\frac{\rmd f}{\rmd r}\frac{\rmd Y_{10}}{\rmd\theta}.
\label{eq:ThetaB}
\end{equation}
We specify the strength of the poloidal field by its magnitude at the poles $r=R_o$, $\theta=0,\pi$: the $\theta$-component vanishes here, and since $Y_{10}(\theta=0)=\sqrt{3/(4\pi)}$, $B_r(r=R_o,\theta=0)=B_p$, so the constant $B_p$ sets the surface dipole field strength.

We will choose as our initial field either a purely dipolar poloidal field as described above, or a mixed poloidal-toroidal field with dipolar poloidal and quadrupolar toroidal components. From Eq.~(\ref{eq:BVectorSpherical}), the quadrupolar toroidal field will have form $B^{(2)}_{2 0}(r)\bm{\Phi}_{2 0}(\theta)$ where $B^{(2)}_{2 0}(r)$ satisfies the boundary condition Eq.~(\ref{eq:OuterBCToroidal}). We also impose that $B^{(2)}_{2 0}(r)$ vanishes at $r=R_i$. We thus take
\begin{equation}
B_{\phi} = 32\sqrt{\frac{4\pi}{45}}B_t\frac{(r-R_i)^2(R_o-r)^2}{(R_o-R_i)^4}\frac{\rmd Y_{20}}{\rmd\theta},
\label{eq:PhiB}
\end{equation}
Eq.~(\ref{eq:PhiB}) is maximized when $r=(R_i+R_o)/2$, and $|\rmd Y_{20}/\rmd \theta|$ takes a maximum value of $\sqrt{45/(16\pi)}$, so $B_t$ is the maximum strength of the initial toroidal field. It is well known in electron MHD that for $\bm{H}=\bm{B}$, if the initial field is toroidal, will remain toroidal, but a poloidal initial field will generate a nonzero toroidal component (e.g.,~\citet{Vainshtein2000,Reisenegger2007,Vigano2012,Kojima2012}). This continues to hold true if $\bm{H}\neq\bm{B}$, since the new terms in the magnetic induction equation for the poloidal component--those involving the magnetization and its derivatives--also depend on the poloidal field or poloidal field direction vector.

\subsection{Thermal evolution}

The thermal evolution of the outer crust of neutron stars presents numerical difficulties, because it spans four-five orders of magnitude in density and will thus contain a large temperature gradient. A common way to handle this problem, which we also employ, is to treat the outer portion of the crust using the quasi-stationary envelope approximation~\citep{Gudmundsson1983}. In this approximation, an outer layer of the crust is assumed to contain no heat sinks or sources such that the heat flux at the high-density end of this layer, denoted with radius $r=R_b$, equals the heat flux escaping through the outer surface of the star, given by the black-body luminosity $\sigma_{\rm SB}T_s^4(T_b)$, where $\sigma_{\rm SB}$ is the Stefan--Boltzmann constant, $T_s$ is the local surface temperature and $T_b$ is the local temperature at radius $R_b$. The temperature $T(r=R_b)=T_b$ and temperature gradient at $r=R_b=R_o$ can be related to the surface temperature, and hence a boundary condition can be imposed here involving only $T_b$.

While we considered the~\citet{Gudmundsson1983} envelope model in benchmarking simulations, it has been known for decades that a strong magnetic field in the crust dramatically modifies the heat blanketing envelope~\citep{Hernquist1985,VanRiper1988,Schaaf1990,Shibanov1996,Potekhin2001,Ventura2001,Potekhin2003,Potekhin2005,Potekhin2007,Pons2009,Beznogov2021,
Dehman2023b}, with the exact changes to the resulting surface temperature distribution depending on the strength and configuration of the surface field. Since we are interested in strong fields, in the simulations featured in this paper we use the updated magnetic field-dependent envelope from Appendix B of~\citet{Potekhin2015}. This model provides a function $T_s=T_s(T_b,\bm{B})$ relating the local surface temperature $T_s$ to $T_b$ and also to the magnetic field strength and configuration at the surface. Its only field dependence is on the strength of the dipole field at the poles. The boundary condition is specified by equating the radial heat flux density at the outer simulation boundary with the blackbody emission at the true surface of the star:
\begin{equation}
\bm{q}(r=R_o)\cdot\hat{\bm{r}}=\left(\frac{R}{R_o}\right)^2\sigma_{\rm{SB}}T_s^4(T_b,\bm{B}).
\label{eq:OuterBCHeatFlux}
\end{equation}
Here $(R/R_o)^2\approx 1$, where $R$ is the outer radius of the star. We use this expression to impose a condition on the temperature in the first ghost cell at the outer radial boundary, thus enforcing that the radial heat flux density at the outer boundary is given by Eq.~(\ref{eq:OuterBCHeatFlux}). We set the upper density $\rho_b$ of the heat-blanketing envelope to $\rho_b=1.5\times10^{10}$ g/cm$^3$.

To correctly model the thermal boundary condition at the crust-core transition, we consider a single boundary ghost cell ``core''~\citep{Ascenzi2024} with uniform redshifted temperature $\tilde{T}_{\rm core}$, an excellent approximation for most of the star's lifetime due to the high thermal conductivity of the core. Dropping the Joule heating term and assuming isothermal conductivity, integration of Eq.~(\ref{eq:GeneralizedHeatEquationStart}) over the volume of the star gives
\begin{equation}
C_V\frac{\rmd\tilde{T}_{\rm core}}{\rmd t} = -\dot{\tilde{Q}}_{\nu} - \oint_{\partial V_{\rm core}}{\rm e}^{\nu}\bm{q}\cdot\rmd\bm{A},
\label{eq:CoreHeatEq}
\end{equation}
where $C_v$ and $\dot{\tilde{Q}}_{\nu}>0$ are the volume-integrated heat capacity and redshifted neutrino luminosity over the entire core volume $V_{\rm core}$:
\begin{align}
C_V={}&4\pi\int_0^{R_i} \rmd r r^2e^{\lambda/2}c_V(T=\rme^{-\nu/2}\tilde{T}_{\rm core}),
\\
\dot{\tilde{Q}}_{\nu}={}&4\pi\int_0^{R_i}\rmd r r^2e^{\lambda/2}\rme^{\nu}\dot{q}_{\nu}(T=\rme^{-\nu/2}\tilde{T}_{\rm core}).
\end{align}
The only heat diffusion appearing in Eq.~(\ref{eq:CoreHeatEq}) is that between the crust and the core at the outer surface of the core, due to the uniform $\tilde{T}$ assumption within the core volume. Like the crust, the core is modelled using the BSk24 EOS, which considers a composition of neutrons, protons, electrons and muons. The heat capacity of the core consists of the free Fermi gas contributions for the leptons calculated using the Sommerfeld approximation, and hence $\propto T$~\citep{Potekhin2015}, and the same form for the nucleons except replacing the bare masses with their Landau effective masses. The neutrino emissivity of the core is taken to be a sum of the modified Urca (mUrca) and direct Urca (dUrca) reactions from both nucleon species and both lepton species~\citep{Yakovlev2001}, nucleon-nucleon bremsstrahlung ($nn$, $np$ and $pp$)~\citep{Yakovlev1995}, and the $^3$P$_2$, $m_J=0$ (``type B'') superfluid neutron Cooper pair breaking and formation (PBF) neutrino emission~\citep{Yakovlev2001}, with the reduction due to collective effects implemented~\citep{Leinson2010}. Superfluid reduction factors for both $^1$S$_0$ proton superconductivity and type B neutron superfluidity for the nucleon specific heat capacities, mUrca neutrino emissivity and dUrca neutrino emissivity are taken from~\citet{Yakovlev1999},~\citet{Gusakov2002} and ~\citet{Yakovlev2001} respectively. The Landau effective masses of the nucleons appearing in the core neutrino emissivities and nucleonic specific heat capacities were computed directly from the Skyrme forces used to generate the BSk22-26 EOS using Eq.~(A10) of~\citet{Chamel2009} and the parameters of Table II of~\citet{Goriely2013}. The in-medium correction factors to the modified Urca emissivity from~\citet{Shternin2018} have also been applied.

The direct Urca threshold mass for the BSk24 EOS is 1.595$M_{\odot}$, and since we focus on studying a 1.4$M_{\odot}$ model, the direct Urca reactions play no role in the simulations in this paper. We compute interpolating functions for $C_V$ and $\dot{Q}_{\nu}$ as a function of temperature Eq.~(\ref{eq:CoreHeatEq}) is solved alongside the magnetothermal evolution for the crust using the same timestepping method as the crust. The interior ghost cell temperatures are set to $\tilde{T}_{\rm core}$, thus allowing heat to flow into/out of the core from the crust.

In all simulations in this paper, we assume a uniform redshifted temperature of $5\times10^{9}$ K throughout the star as the initial condition. As argued in e.g.,~\citet{Potekhin2018}, and as we also find, the results are insensitive to the initial temperature choice for any sufficiently high temperature $\gtrsim 10^9$ K, as the initial temperature profiles evolve to become nearly identical within timescales of less than one day through rapid neutrino cooling.

\subsection{Additional approximations}
\label{sec:AdditionalApproximations}

As discussed in RW25 and earlier papers~\citep{Blandford1982,Suh2010,Wang2013,Wang2016,Rau2023}, there are regions of the $(B, \mue, T)$ parameter space for which the magnetized free electron gas is thermodynamically unstable to magnetic domain formation. This occurs when $\chi_{\mu}>1/(4\pi)$, and in equilibrium in these unstable regions of parameter space $\chi_{\mu}$ will take a value slightly less than $1/(4\pi)$. Like in RW25, we assume that magnetic domains form on timescales much shorter than the Hall timescale for the evolution of the large-scale magnetic field, and we use the approximate criterion that domains will form where $\chi_{\mu}>1/(4\pi)$ and set $\chi_{\mu}=1/(4\pi)$ when this condition is true such that $\partial^2\Omega_{\rme}/\partial B^2=(1/(4\pi)-\chi_{\mu})\geq 0$ and thermodynamic stability is always maintained. 

\section{The \texttt{QMFM} code}
\label{sec:QMFM}

To perform the magnetothermal evolution simulations, we wrote a 2.5D finite volume code in axisymmetric spherical coordinates i.e., three magnetic field components with dependence on two spatial coordinates $r$ and $\theta$. The main distinguishing factor between this code and other finite volume codes~\citep{Vigano2012,Vigano2013,Vigano2021,Dehman2023,Ascenzi2024} used to solve the neutron star magnetothermal evolution code is the inclusion of Landau quantization effects on all transport and thermodynamic properties and the inclusion of the Landau quantization-induced magnetization field, hence the code's name, Quantizing Magnetic Fields in Magnetars (\texttt{QMFM}). \texttt{QMFM} is written in C++ and MPI-parallelized by domain decomposition.

The magnetic field evolution in \texttt{QMFM} is based on the constrained transport method~\citep{Evans1988} as expounded for axisymmetric electron MHD in~\citet{Vigano2012,Vigano2021}, which guarantees $\bm{\nabla}\cdot\bm{B}=0$ to within machine precision throughout the evolution as long as the initial magnetic field is divergence-free. We evolve the toroidal field $B_{\phi}$ differently than the poloidal field to properly solve for its Burgers equation-like behavior. The flux is reconstructed at the upwind interface using the monotonized central-difference (MC) limiter. We write the magnetic induction equation for $B_{\phi}$ in a slightly different form than Eq.~(14) of~\citet{Vigano2021}, and also incorporate the novel magnetization terms. We thus evolve $B_{\phi}$ using
\begin{align}
\frac{\partial B_{\phi}}{\partial t} ={}& -\frac{\rme^{-\lambda/2}}{r}\frac{\partial}{\partial r}\left(r\frac{\beta_rB_{\phi}^2}{2}\right)-\frac{1}{r}\frac{\partial}{\partial\theta}\left(\frac{\beta_{\theta}B_{\phi}^2}{2}\right)
\nonumber
\\
{}&-\frac{\rme^{-\lambda/2}}{r}\frac{\partial}{\partial r}\left({\rm e}^{\nu/2}r\frac{J_{\phi}B_r}{en_{\rm e}}\right)-\frac{1}{r}\frac{\partial}{\partial\theta}\left({\rm e}^{\nu/2}\frac{J_{\phi}B_{\theta}}{en_{\rm e}}\right)
\nonumber
\\
{}&+\frac{1}{r}\frac{\partial}{\partial\theta}\left(c\rho_{\rm \parallel}J_{\parallel,r}\right)-\frac{{\rm e}^{-\lambda/2}}{r}\frac{\partial}{\partial r}\left(rc\rho_{\rm \parallel}J_{\parallel,\theta}\right)
\nonumber
\\
{}&+\frac{1}{r}\frac{\partial}{\partial\theta}\left(c\rho_{\rm \perp}J_{\perp,r}\right)-\frac{{\rm e}^{-\lambda/2}}{r}\frac{\partial}{\partial r}\left(rc\rho_{\rm \perp}J_{\perp,\theta}\right),
\end{align}
where
\begin{align}
\beta_r ={}& \rme^{\nu/2}\frac{\sin^2\theta}{r}\frac{\partial}{\partial\theta}\left(\frac{\overline{\eta}_{\rm H}}{\sin^2\theta}\right)
\nonumber
\\
{}&+ \frac{8\pi}{rB}\left[ {\rm e}^{\nu/2}\hat{\eta}_{\rm H}\frac{\partial B}{\partial \theta} + {\rm e}^{\nu/2}\eta_{\rm H}\mathcal{M}_{\mu}\frac{\partial\mue}{\partial \theta} + \eta_{\rm H}\mathcal{M}_T\frac{\partial\tilde{T}}{\partial \theta}\right],
\label{eq:beta_r}
\\
\beta_{\theta} ={}& -\rme^{\nu-\lambda/2}r^2\frac{\partial}{\partial r}\left(\frac{\overline{\eta}_{\rm H}}{r^2}\rme^{-\nu/2}\right) 
\nonumber
\\
{}&- \frac{8\pi}{B}\left[ {\rm e}^{\nu/2}\hat{\eta}_{\rm H}\frac{\partial B}{\partial r} + {\rm e}^{\nu/2}\eta_{\rm H}\mathcal{M}_{\mu}\frac{\partial\mue}{\partial r} + \eta_{\rm H}\mathcal{M}_T\frac{\partial\tilde{T}}{\partial r}\right],
\label{eq:beta_theta}
\\
\overline{\eta}_{\rm H} \equiv{}& \left(1-4\pi\frac{M}{B}\right)\eta_{\rm H}, \qquad \hat{\eta}_{\rm H}\equiv\left(\chi_{\mu}-\frac{M}{B}\right)\eta_{\rm H}.
\label{eq:eta_H_mod}
\end{align}
$\eta_{\rm H}=c/(4\pi en_{\rm e})$ is the Hall diffusivity divided by $B$. Eq.~(\ref{eq:beta_r}--\ref{eq:beta_theta}) determine the advection speeds in the radial and polar directions $\beta_rB_{\phi}$ and $\beta_{\theta}B_{\phi}$, which determine the upwind direction. This definition of $\beta_r$ also fully accounts for variation of $\overline{\eta}_{\rm H}$ in the $\theta$-direction, which enters through its dependence on $B$ and $M$. Even in the case ignoring magnetization, $\beta_r$ is nonzero in spherical coordinates due to spatial curvature. 

Imposing the potential boundary condition Eq.~(\ref{eq:B1-BrExterior}) requires a Legendre function transformation. We used the \texttt{SHTns} library~\citep{Schaeffer2013} for this purpose.

The timestepping in \texttt{QMFM} is performed using the method of lines: we have so far implemented first and second-order strong stability-preserving (SSP) explicit Runge--Kutta methods~\citep{Gottlieb2001} and an implicit-explicit (IMEX) method, IMEX-SSP2~\citep{Ascher1997,Pareschi2005}. To maintain stability, the timestep for the explicit methods are limited by the Courant--Friedrichs--Lewy (CFL) condition. For thermal and magnetic (Hall) evolution respectively the CFL-limited timestep is 
\begin{align}
(\Delta t)_T {}&\leq k_{{\rm C},T}{\rm min}\left(\frac{\Delta L^2}{\eta_T}\right),
\\
(\Delta t)_B {}&\leq k_{{\rm C},B}{\rm min}\left(\frac{\Delta L^2}{(1-4\pi\chi_{\mu})(B\eta_{\rm H}+\eta_{\rm O})})\right).
\end{align}
Here $\eta_T = \kappa/c_V$ is the thermal diffusivity, $\eta_{\rm O}=\rho_{\perp}c^2/4\pi$ is the Ohmic diffusivity and $(1-4\pi\chi_{\mu})$ is the approximate amplification of $\eta_{\rm H}$ and $\eta_{\rm O}$ due to the Landau quantization-induced magnetic susceptibility. $k_{{\rm C},T}$ and $k_{{\rm C},B}$ are the Courant numbers for the thermal and magnetic evolution, $\Delta L$ is the spatial grid size and ${\rm max}(f)$ is the maximum value of $f$ across the simulation domain. $k_{{\rm C},T}$ and $k_{{\rm C},B}$ are both less than unity, though the maximum value they can take is difficult to determine theoretically in problems involving non-uniform diffusivities and stiff source terms.

We exclusively run the simulations discussed in this paper using the IMEX-SSP2 method, treating the thermal evolution implicitly and the magnetic evolution explicitly. This is because $(\Delta t)_T$ is one to two orders of magnitude smaller than $(\Delta t)_B$ for the magnetic field and temperature ranges of interest, and running the simulations using $(\Delta t)_T$ as the timestep would be prohibitvely slow. Because the heat equation contains terms which are nonlinear in $T$ due to the stiff neutrino emissivity term and the temperature-dependence of the thermal conductivity, we do not expect our implicit evolution of the temperature to be unconditionally stable. Regardless, the maximum timestep is CFL-limited by the magnetic evolution evolution and so cannot be taken arbitrarily large. We thus set the timestep equal to $(\Delta t)_B$, except at the beginning of the simulations, where we consider a gradual ramp-up to $(\Delta t)_B$ to avoid overly damping the initial rapid neutrino emission in the heat equation. We set $k_{{\rm C},B}=0.2$, and find that this choice of timestep results in stable magnetic and thermal evolution.

The implicit part of the IMEX timestep uses an alternating direction-implicit (ADI) scheme for mixed derivatives~\citep{Craig1988}, which is necessary for the case of anisotropic thermal conductivity. Use of an ADI scheme maintains the tridiagonal nature of the matrix to be inverted in the implicit step, allowing us to use the Thomas algorithm to efficiently compute the matrix-vector product to evolve $T$. At the interior boundaries between partial domains, we use an explicit method to guess the temperature of the ghost cells at the next timestep and use this as a boundary value within the implicit timestep for the temperature in the non-ghost cells. We then exchange the temperature in the boundary cells between parallelized domains and iterate the implicit solver until the solution converges. We find that the solution converges within a few iterations using this method. 

\section{Simulations}
\label{sec:SimResults}

To benchmark \texttt{QMFM}, we first ran magnetothermal simulations without including the novel Landau quantization effects to compare to previous magnetic and magnetothermal evolution studies of neutron star crusts~\citep{Vigano2013,Gourgouliatos2014,Dehman2023a}. The main product of these simulations which can be compared to observation are thermal luminosity curves, which are computed in various references for magnetothermal simulations~\citep{Vigano2013,Dehman2023b,Ascenzi2024,Suvorov2026}, and for thermal only simulations with magnetic field effects partially included in~\citet{Potekhin2018}.

The dHvA oscillations of $M$, $\chi_{\mu}$, $\mathcal{M}_{\mu}$ and $\mathcal{M}_T$ can have density and $B$-dependent spatial periods which vary on length scales $\lesssim 1$ m. This is far shorter than what we can reasonably resolve while simultaneously running the simulations for timescales approaching the typical ages of magnetars $\lesssim 10^5$ yr. We are thus unable to completely resolve the dHvA oscillations, but even if they are not completely resolved, they will still affect the simulations, with the lower resolution resulting in coarse-grained ``averages'' of $M$, $\chi_{\mu}$, $\mathcal{M}_{\mu}$ and $\mathcal{M}_T$ appearing in the magnetothermal evolution equations. %For this reason, we perform simulations with Landau quantization effects enabled at two different spatial resolutions to estimate the effect of this coarse-grained averaging over the dHvA oscillations on the long-term magnetothermal evolution.

The series of simulations that we consider here are described in Table~\ref{tab:Simulations}. These include three different initial field configurations--two purely poloidal with different initial strengths $10^{14}$ and $2\times10^{15}$ G ($B1$ and $B3$), and a mixed poloidal-toroidal field with initial poloidal and toroidal field strengths $5\times10^{14}$ and $10^{15}$ G respectively ($B2$). They also include simulations with and without Landau quantization effects enabled, at different resolutions to determine the effect of resolution on the Landau quantization-induced amplification of the Ohmic dissipation, and simulations in the absence of crustal impurities.

\begin{deluxetable}{lcccccc}
\tablecaption{Summary of the simulations considered in this paper. $N_r$ and $N_{\theta}$ are the resolutions in the radial and $\theta$-directions, $B_p$ and $B_t$ are the initial strengths of the poloidal and toroidal fields (see Section~\ref{sec:InitialMagneticField} for definitions) and whether Landau quantization effects (LQ) and impurities (Imp.) are enabled (Y) or disabled (N).\label{tab:Simulations}}
\tablehead{
\colhead{Name} &
\colhead{$N_r$} &
\colhead{$N_{\theta}$} &
\colhead{$B_p$ (G)} &
\colhead{$B_t$ (G)} &
\colhead{LQ} &
\colhead{Imp.}
}
\startdata
$B1_{\rm NLQ}$    & 40 & 60 & $10^{14}$        & 0          & N & Y \\
$B2_{\rm NLQ}$    & 40 & 60 & $5\times10^{14}$ & $10^{15}$  & N & Y \\
$B3_{\rm NLQ}$    & 60 & 60 & $2\times10^{15}$ & 0          & N & Y \\
$B1_{\rm LQ}$     & 40 & 60 & $10^{14}$        & 0          & Y & Y \\
$B2_{\rm LQ}$     & 40 & 60 & $5\times10^{14}$ & $10^{15}$  & Y & Y \\
$B3_{\rm LQ}$     & 60 & 60 & $2\times10^{15}$ & 0          & Y & Y \\
$B1_{\rm LQ2}$    & 80 & 120 & $10^{14}$       & 0          & Y & Y \\
$B2_{\rm LQ2}$    & 80 & 120 & $5\times10^{14}$        & $10^{15}$  & Y & Y \\
$B1_{\rm LQ,NI}$  & 40 & 60 & $10^{14}$        & 0          & Y & N \\
$B2_{\rm LQ,NI}$  & 40 & 60 & $5\times10^{14}$ & $10^{15}$  & Y & N \\
\enddata
\end{deluxetable}

\subsection{Magnetothermal evolution without Landau quantization}

\begin{figure*}
\centering
\includegraphics[width=0.97\textwidth]{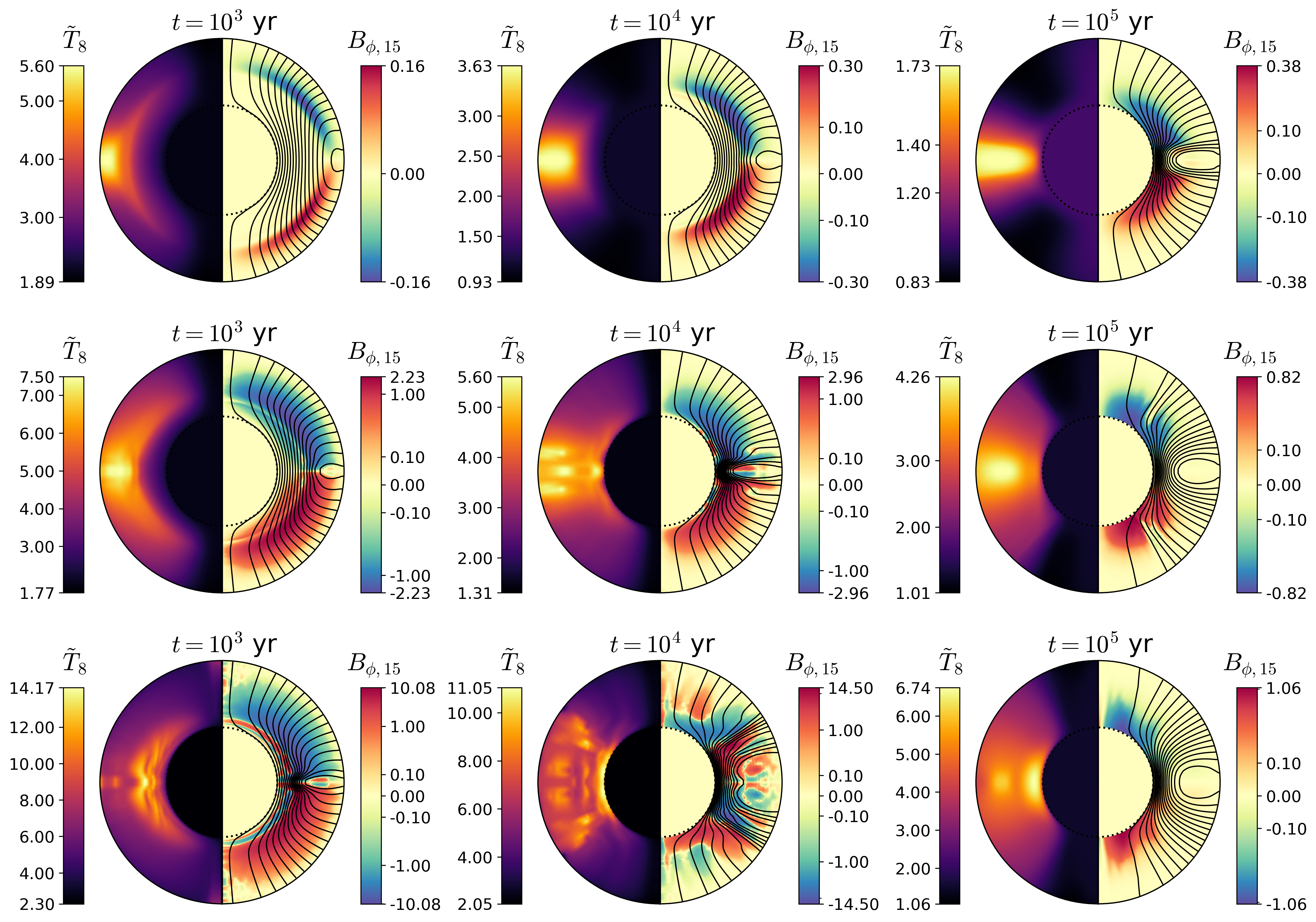}
\caption{Internal redshifted temperature (left) and magnetic field (right) for the simulations without Landau quantization $B1_{\rm NLQ}$ (top row), $B2_{\rm NLQ}$ (middle row), $B3_{\rm NLQ}$ (bottom row) at three different times $t=10^3$, $10^4$ and $10^5$ yr. The temperature is given in units of $10^8$ K, and the toroidal magnetic field $B_{\phi}$ in units of $10^{15}$ G. The poloidal field lines are shown as solid black lines on the right, and the crust-core transition is shown as a dotted black line. The crust thickness has been exaggerated by a factor of eight  and the color bar scale for $B_{\phi}$ is logarithmic in $|B_{\phi}|$ for improved visualization.}
\label{fig:MagnetothermalSimulationsNoLQ}
\end{figure*}

To compare with previous magnetothermal simulations, we first performed a series of simulations without including the effects of Landau quantization: $B1_{\rm NLQ}$, $B2_{\rm NLQ}$, $B3_{\rm NLQ}$ from Table~\ref{tab:Simulations}. Figure~\ref{fig:MagnetothermalSimulationsNoLQ} shows snapshots of the internal temperature and magnetic field for each of these models at three different times. 

In agreement with previous studies, we observe common aspects of Hall evolution: the generation of a quadrupolar toroidal field if in the absence of an initial toroidal field or its enhancement if a toroidal field is already present, the advection of the toroidal field toward the crust-core boundary, the generation of higher-order poloidal multipoles by the quadrupolar toroidal field, and the subsequent generation of higher-order toroidal multipoles in a Hall cascade. The poloidal field lines are pulled towards the crust-core boundary and the equator, leading to a larger equatorial region where heat can be trapped at later times. The rate at which this all takes place increases linearly with the field strength, and so the advection of the toroidal field to the crust-core boundary, the equatorial pinching of the poloidal field, and the generation of higher-order multipoles are more developed within simulation $B3_{\rm NLQ}$ than $B2_{\rm NLQ}$ or $B1_{\rm NLQ}$.

In terms of the thermal evolution, the crust and core both rapidly cool below $\tilde{T}=10^9$ K by neutrino emission, with the core initially cooling faster than the crust. Heat conduction in the crust along the poloidal field lines to the poles and subsequent radiation from the surface quickly leaves the polar regions at a lower temperature than the equator. An equatorial hot spot is maintained by strong Joule heating from the equatorial region, with the heat being trapped here by the strong poloidal magnetic field. The temperature of the hot spot is set by the balance of the Joule heating and the neutrino cooling/surface photon emission in this region: the configurations with stronger magnetic fields more quickly generate strong currents in the equatorial region and greater heating, and so their hot spots reach higher temperatures. Impurities in the inner crust are important in increasing the Joule heating, and in their absence, the heat generation in the equatorial region will be insufficient to prevent most of the inner crust from dropping below the neutron superfluid critical temperature, resulting in a greatly decreased heat capacity and the equatorial region becoming cooler than the poles. These results are broadly similar to those found in previous studies, with differences that can be attributed to different equations of state and minor differences in microphysics. 

\begin{figure}
\centering
\includegraphics[width=0.98\linewidth]{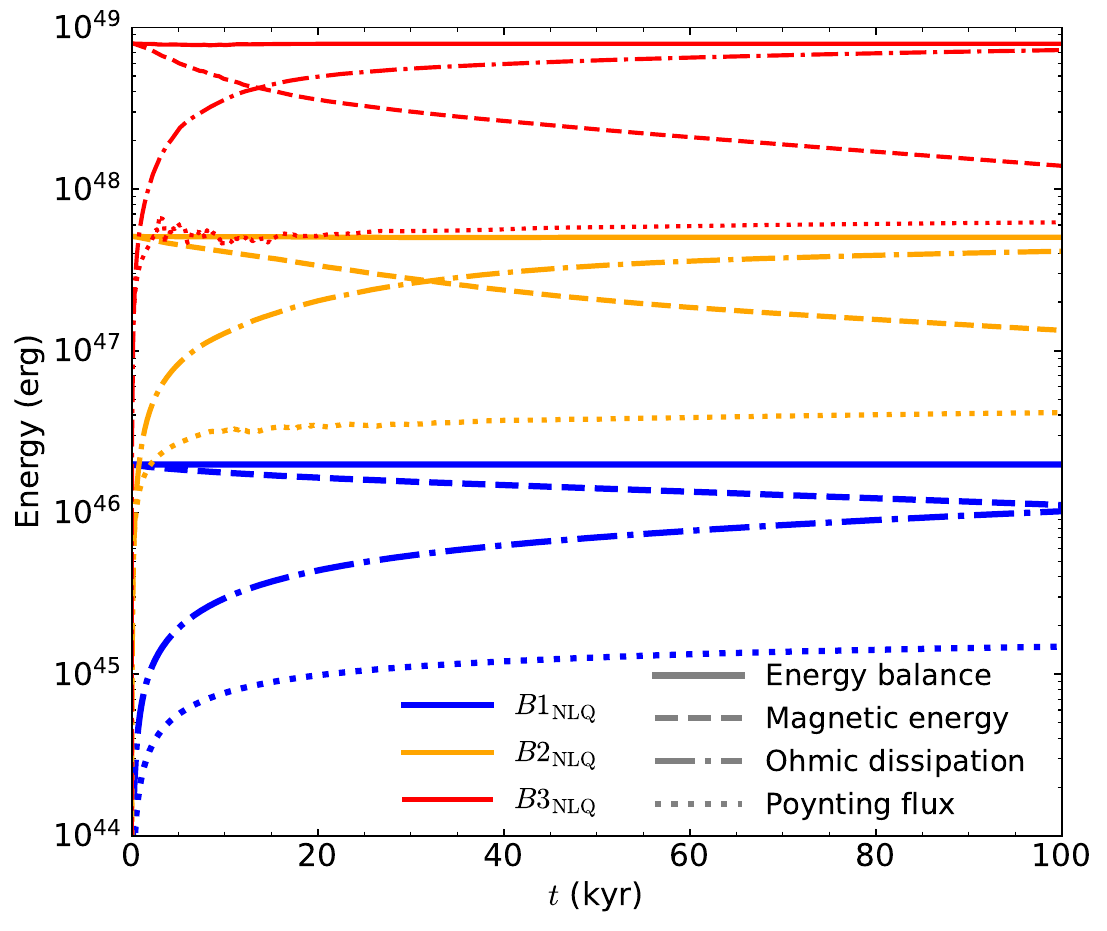}
\caption{Global energy conservation for the simulations $B1_{\rm NLQ}$, $B2_{\rm NLQ}$ and $B3_{\rm NLQ}$, corresponding to the blue, red and orange lines. The total energy balance (solid line) is a sum of the magnetic energy (dashed line) plus the total energy that is Ohmically dissipated (dash-dotted line) and the Poynting flux losses through the outer surface (dotted line).}
\label{fig:EnergyConservation}
\end{figure}

Figure~\ref{fig:EnergyConservation} shows the global energy conservation for these simulations as computed by integrating Eq.~(\ref{eq:LocalEnergyConservation}) (with $\bm{B}=\bm{H}$) over the simulation volume. In this case, the energy balance should equal the magnetic energy plus the Ohmic dissipation and the Poynting flux losses. The energy is conserved to within $\sim3$\% for each of the simulations. For strong poloidal fields, initially the energy within the simulation domain increases slightly due to an inward-directed Poynting flux at the outer surface, corresponding to magnetic field energy from the exterior potential field, imposed as the outer boundary condition. While the magnitude of the energy increase from this Poynting flux is comparable to the Ohmic dissipation at early times $\lesssim$ a kyr, it is an order of magnitude or more smaller after $\sim$ 10 kyr.

\subsection{Magnetothermal evolution with Landau quantization of electrons}

\begin{figure*}
\centering
\includegraphics[width=0.97\textwidth]{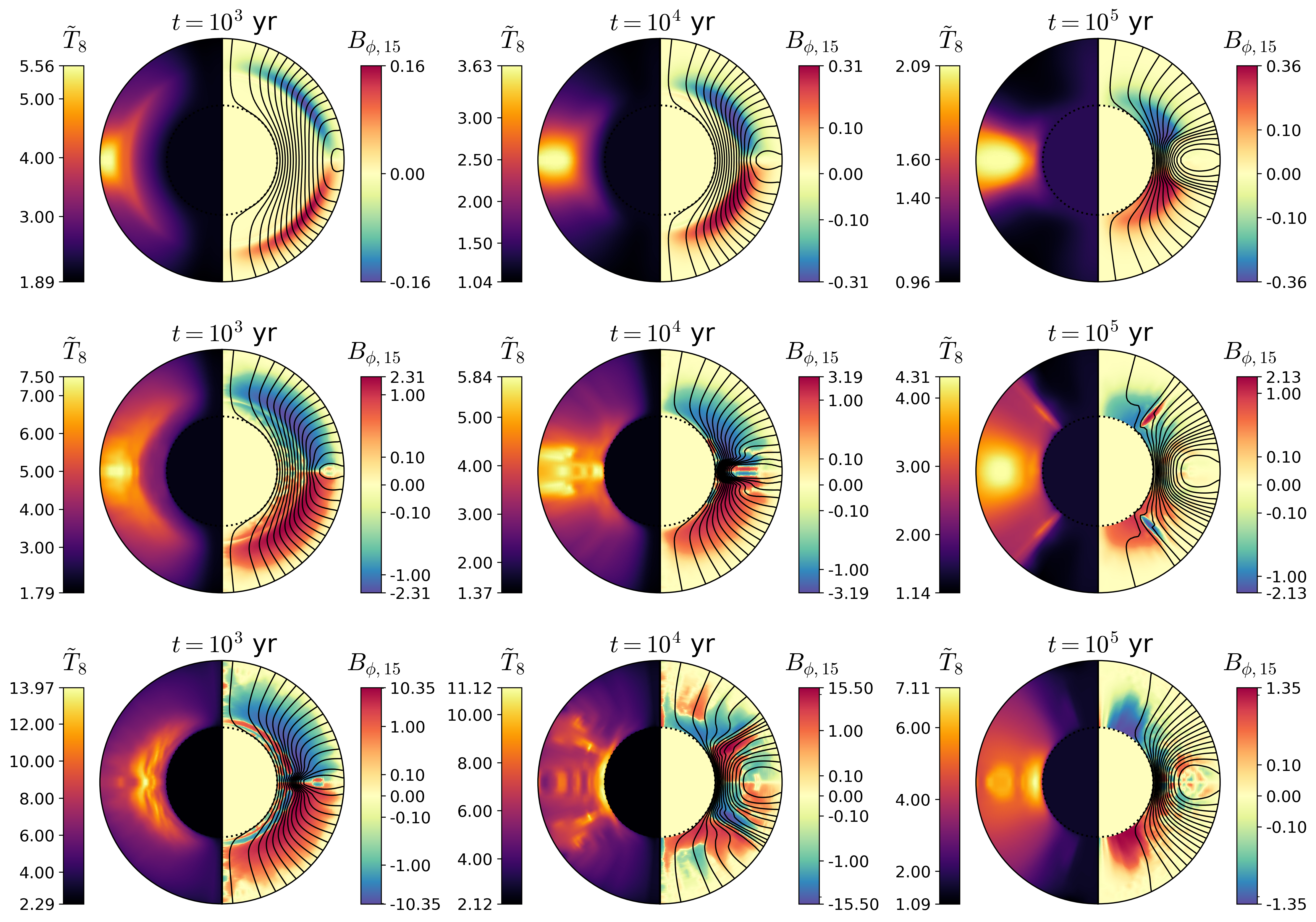}
\caption{Same as Figure~\ref{fig:MagnetothermalSimulationsNoLQ} except for the simulations with Landau quantization $B1_{\rm LQ}$ (top row), $B2_{\rm LQ}$ (middle row), $B3_{\rm LQ}$ (bottom row).}
\label{fig:MagnetothermalSimulationsLQ}
\end{figure*}

\begin{figure}
\centering
\includegraphics[width=0.98\linewidth]{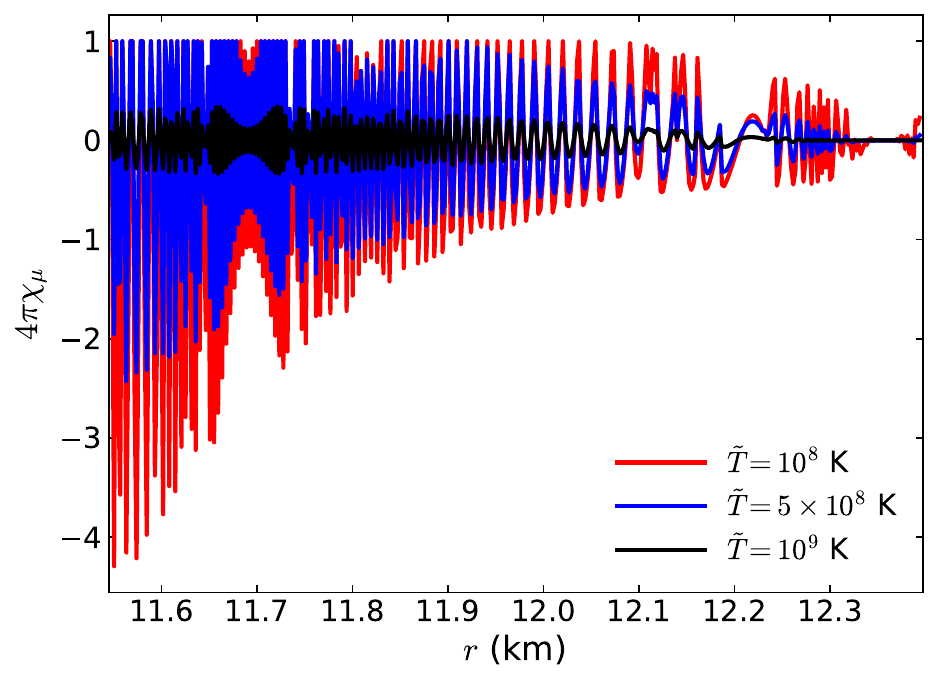}
\caption{The differential magnetic susceptibility $\chi_{\mu}$ multiplied by $4\pi$ within the model neutron star crust at $\theta=\pi/2$ for magnetic field configuration $B2$ and three different redshifted temperatures $\tilde{T}=3\times10^8$, $5\times10^8$, $10^9$ K. Note that much higher radial resolution $N_r$ is used for this plot than is used in the simulations in this paper.}
\label{fig:ChiPlot}
\end{figure}

To study the effects of Landau quantization on the magnetothermal evolution, we perform a direct comparison between otherwise identical simulations with and without it enabled. Simulations $B1_{\rm LQ}$, $B2_{\rm LQ}$ and $B3_{\rm LQ}$ are identical to simulations $B1_{\rm LNQ}$, $B2_{\rm NLQ}$ and $B3_{\rm NLQ}$ except for the inclusion of Landau quantization. Figure~\ref{fig:MagnetothermalSimulationsLQ} shows snapshots of the internal temperature and magnetic field for these simulations. Despite the inclusion of the Landau quantization, the results are broadly similar to those of Figure~\ref{fig:MagnetothermalSimulationsLQ}, especially at $t=1$ and $10$ kyr. The most notable difference at these times is the generation of additional small-scale magnetic field structures (i.e., higher-order multipoles) by the Landau quantization-induced dHvA oscillations. This is most prominent near the crust-core boundary, where the magnetic field is strongest and thus the amplitude of the dHvA oscillations is generally largest, since for stronger fields the thermal suppression of the oscillations is reduced. This is illustrated in Figure~\ref{fig:ChiPlot}, which shows the dHvA oscillations of $\chi_{\mu}$, the quantity most responsible for the enhanced Ohmic dissipation, within the crust for field configuration $B2$ at three different temperatures. The amplitude of these oscillations is largest near the inner radius $R_i$, and can be nearly eight times as large as the amplitude near the surface at $\tilde{T}=3\times10^8$ K. 

At later times, the effect of the Landau quantization is most pronounced. This is both because its effect has accumulated over the entire simulation, and also because the crust has further cooled, allowing the dHvA oscillations to grow to larger amplitude. The crust is generally hotter, and more prominent small-scale field features have been generated. The latter effect is less noticeable for $B1_{\rm LQ}$, but the equatorial hotspot is still $\sim20$\% hotter after $100$ kyr simulation than in $B1_{\rm NLQ}$. The small-scale features of the toroidal field in $B2_{\rm LQ}$ and $B3_{\rm LQ}$ are significantly stronger than in their respective Landau quantization-free counterparts. The equatorial hotspot at later times is also at a higher temperature in these two simulations than in $B2_{\rm NLQ}$ and $B3_{\rm NLQ}$, and although this is only marginally so in $B2_{\rm LQ}$ compared to $B2_{\rm NLQ}$, the crust is on average significantly warmer in $B2_{\rm LQ}$, with the heat being more uniformly distributed than concentrated in the equatorial region.

\begin{figure*}
\centering
\includegraphics[width=0.97\textwidth]{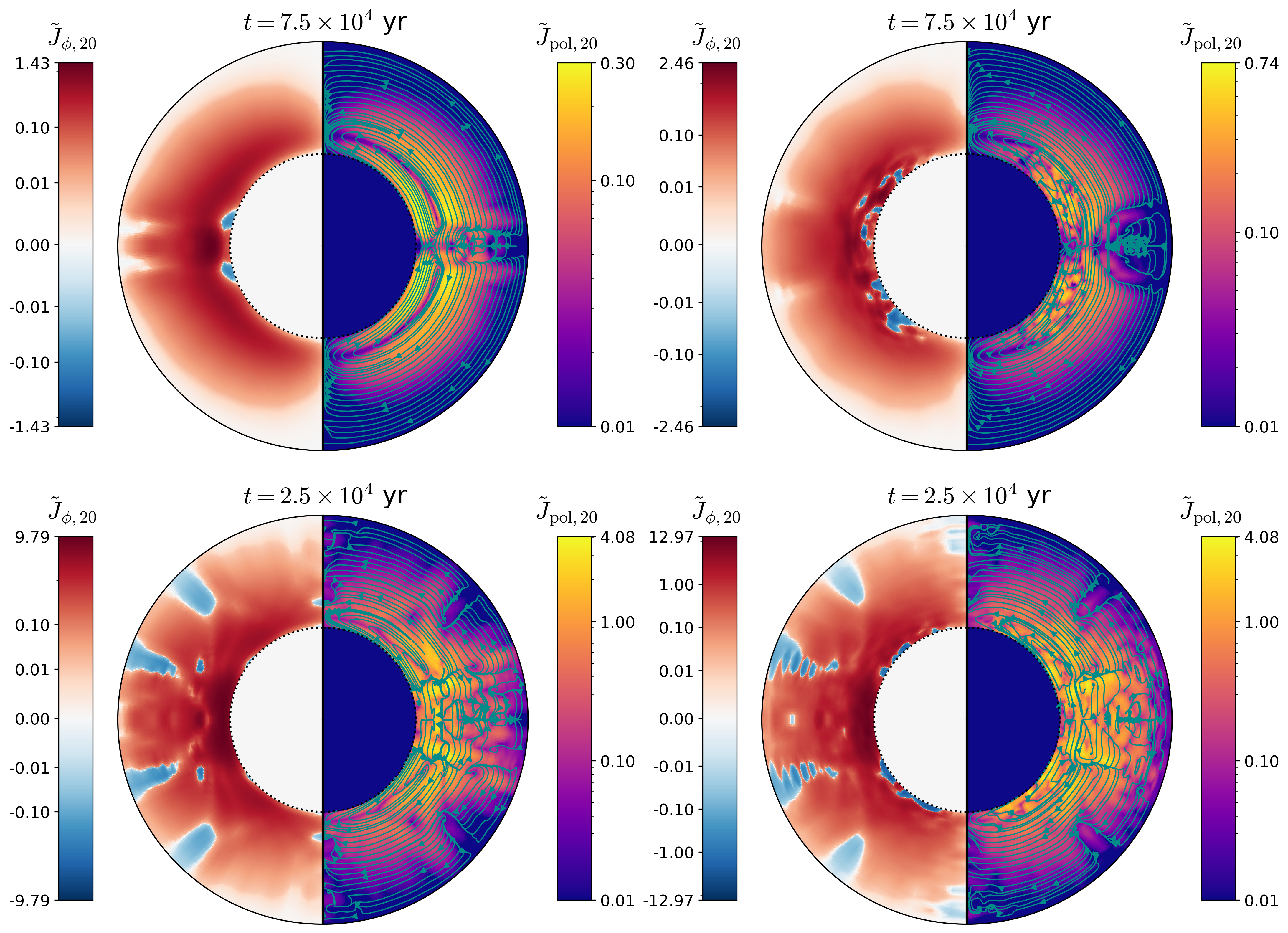}
\caption{Toroidal current density $\tilde{J}_{\phi}=\rme^{\nu/2}J_{\phi}$ (left) and magnitude of the redshifted poloidal current density $\tilde{J}_{\rm pol}=\rme^{\nu/2}J_{\rm pol}$ (right) in units of $10^{20}$ G/s and at time $t=7.5\times10^4$ (top) or $2.5\times10^4$ yr (bottom). Stream lines of the poloidal current are overlaid on the right. From left to right, top to bottom, results from simulations $B1_{\rm NLQ}$, $B1_{\rm LQ}$, $B2_{\rm NLQ}$ and $B2_{\rm LQ}$ are shown. Note the more complicated current distribution in the two cases with Landau quantization enabled, especially in the inner crust where temperatures are lower and the magnetic field is stronger.}
\label{fig:CurrentDistribution}
\end{figure*}

The difference between the simulations with and without Landau quantization is also seen in the current density, as the dHvA oscillations of $M$, $\chi_{\mu}$, etc. directly enter Eq.~(\ref{eq:JMagnetization}). Figure~\ref{fig:CurrentDistribution} compares the current density inside the star at identical times for simulations which differ only by the inclusion of Landau quantization. The small-scale current features generated by the dHvA oscillations of $M$, $\chi_{\mu}$, etc. are clearly seen in the simulations with Landau quantization, especially in the inner crust. The magnetic field is stronger in this region for our initial magnetic field choices and hence there are more dHvA oscillations here, and as can be seen from Figures~\ref{fig:MagnetothermalSimulationsNoLQ} and~\ref{fig:MagnetothermalSimulationsLQ}, the inner crust is also cooler than the outer crust, where the equatorial hot spot is centered. We also see tmall equatorial asymmetries develop in the current density. With initial conditions such that all quantities are symmetric ($B_{\theta}$, $T$) or antisymmetric ($B_r$, $B_{\phi}$) across the equatorial plane, the simulation should preserve this up to floating point accuracy. We find this to be true for the non-quantizing simulations, but deviations from this quickly appear in the simulations with Landau quantization enabled. The reason for this is the highly peaked thermodynamic functions including the dHvA oscillations, which depend on the maximum number of occupied Landau levels $n_{\rm max}=\lfloor p_F^2/(2eB)\rfloor$ at $T=0$. Since this is a discontinuous function of $B$, calculations involving it will amplify floating point round-off errors that differ across the equatorial symmetry plane. As exact axisymmetric magnetic fields will not be found in neutron stars in the universe, we are fine accepting these small imperfections in our simulations. 

\begin{figure}
\centering
\includegraphics[width=0.98\linewidth]{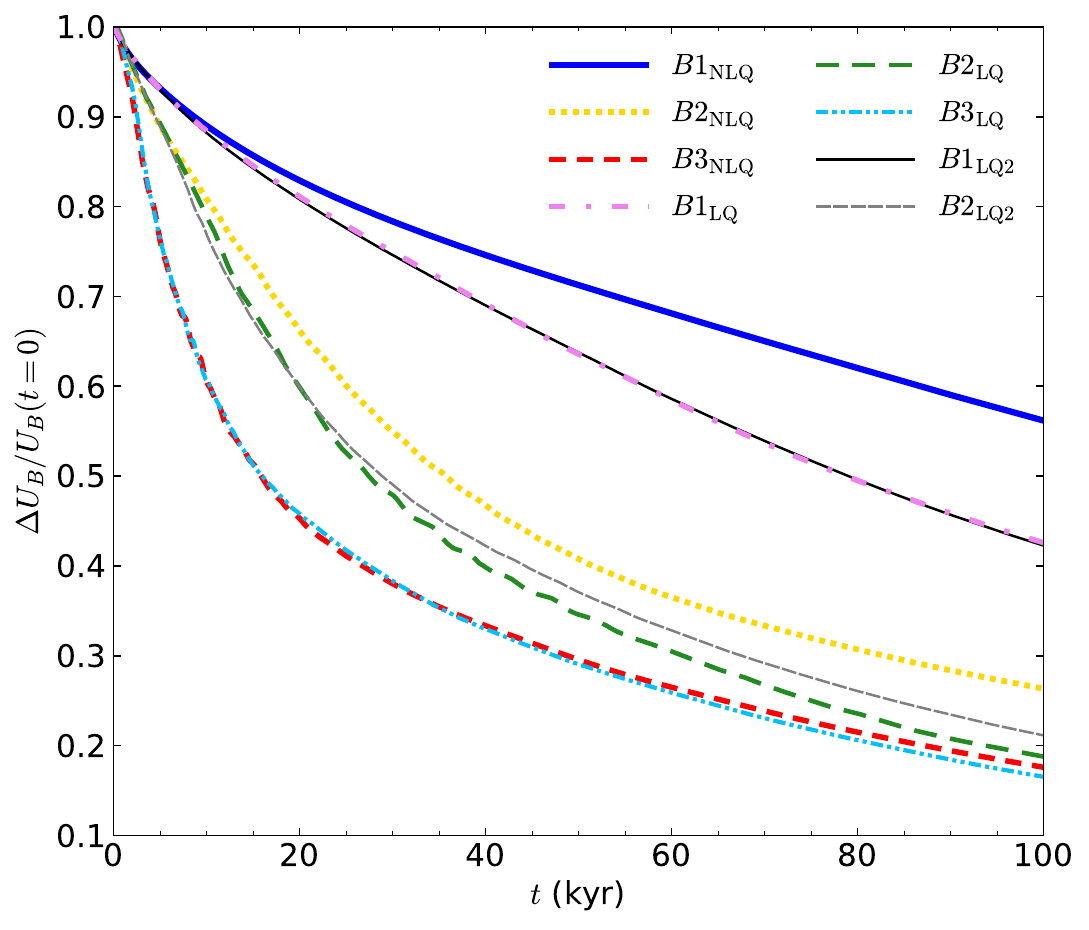}
\caption{Change in magnetic field energy $\Delta U_B=U_B(t)-U_B(t=0)$ over time normalized by the initial magnetic field energy $U_B(t=0)$ for simulations $B1_{\rm NLQ}$ to $B2_{\rm LQ2}$. Simulations $B1_{\rm LQ2}$ and $B2_{\rm LQ2}$ differ from $B1_{\rm LQ}$ and $B2_{\rm LQ}$ only by increased spatial resolution.}
\label{fig:OhmicDissipation}
\end{figure}

The rate of magnetic energy dissipation between the simulations with and without Landau quantization enabled is compared in Figure~\ref{fig:OhmicDissipation}, which shows the change in volume-integrated magnetic field energy $\Delta U_B = U_B(t)-U_B(t=0)$ as a fraction of its initial value. Since the Hall effect generates higher-order, more rapidly-dissipated multipole field components, the stronger the initial field and hence the shorter the Hall time, the faster the field dissipates. We also see that the Landau quantization increases the amount of field energy dissipated within 100 kyr for all three initial field configurations, with the percentage increase in energy dissipated decreasing as a function of field strength from a 30\% increase for configuration $B1$, a 10\% increase for $B2$, to a $\sim1$\% increase for $B3$. As Figures~\ref{fig:MagnetothermalSimulationsNoLQ} and~\ref{fig:MagnetothermalSimulationsLQ} suggest, Landau quantization's effect on the evolution is most pronounced when the fields are weaker but still sufficiently strong to quantize the electrons into a small to moderate number of Landau levels, such that the resulting dHvA oscillations of $M$, $\chi_{\mu}$, etc., are not entirely thermally suppressed. Thus the enhancement tends to be larger for the $10^{14}$ G poloidal field, compared to the stronger initial fields, because the amount of additional Joule heating they provide limits the amount of enhancement to the Joule heating Landau quantization can provide by maintaining the crust at a higher temperature.

%\subsection{Resolution dependence of enhanced Ohmic dissipation}

As RW25 made clear, the enhancement of Ohmic dissipation due to the Landau quantization is a resolution-dependent effect; Figure~\ref{fig:ChiPlot} makes clear that all of the oscillations cannot be resolved in any reasonable electron MHD simulation. Figure~\ref{fig:OhmicDissipation} shows the results of the two high-resolution simulations $B1_{\rm LQ2}$ and $B2_{\rm LQ2}$, identical to $B1_{\rm LQ}$ and $B2_{\rm LQ}$ besides for the increased spatial resolution. We find that the increased resolution has a negligible effect on the energy dissipated for the weaker initial field case $B1_{\rm LQ}$ and $B1_{\rm LQ2}$, and a non-negligible but small effect on the intermediate field case $B2_{\rm LQ}$ and $B2_{\rm LQ2}$, with $2$\% less energy dissipated in $B2_{\rm LQ2}$ than $B2_{\rm LQ}$. This suggests that the lower spatial resolutions used for most of our simulations are still giving a reliable picture of how the Landau quantization effects we include are changing the magnetothermal evolution.

\subsection{Surface photon and neutrino luminosity}

\begin{figure}
\centering
\includegraphics[width=0.98\linewidth]{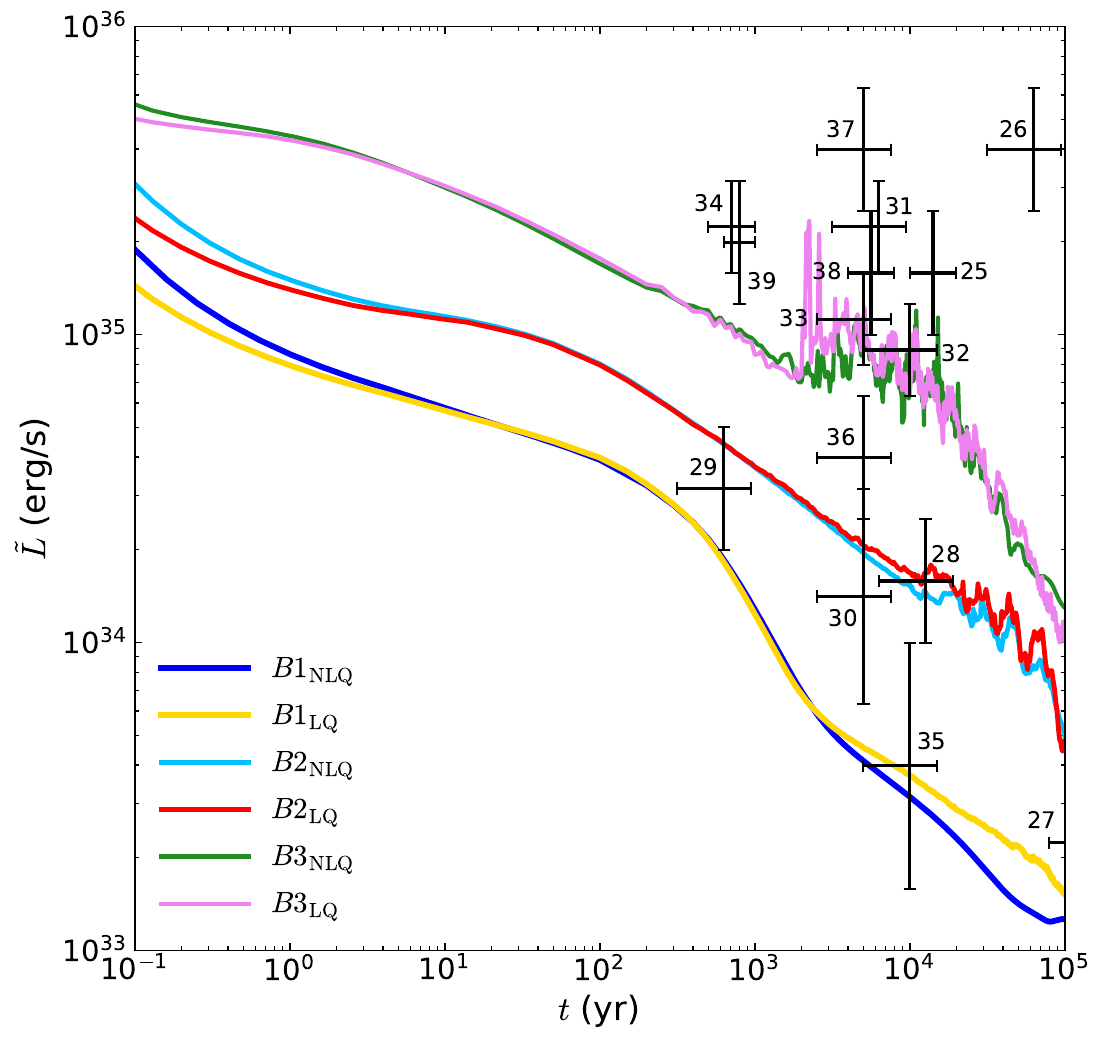}
\caption{Redshifted surface photon luminosity $\tilde{L}$ as a function of time $t$ for four simulations defined in Table~\ref{tab:Simulations}, compared to observational data with error bars for known magnetars from~\citet{Vigano2013} and updated in~\citet{Potekhin2018}, numbered according to the latter. Not all of the 17 magnetars for which we have observational data fit within the bounds of this plot.}
\label{fig:SurfaceLuminosities}
\end{figure}

Comparing the internal temperatures in Figures~\ref{fig:MagnetothermalSimulationsLQ} and~\ref{fig:MagnetothermalSimulationsLQ}, the Landau quantization terms are indeed enhancing the Ohmic dissipation of the magnetic field and keeping the neutron star crust hotter. To determine the extent to which this is reflected in the surface luminosity of the star, we calculate the redshifted surface photon luminosity $\tilde{L}=\rme^{\nu(R)}L_R$ as a function of time, computed by integrating Eq.~(\ref{eq:OuterBCHeatFlux}) over the stellar surface. This is shown in Figure~\ref{fig:SurfaceLuminosities} for six of the simulations, three with and three without Landau quantization enabled. For comparison, we show observational data with error bars for the 17 magnetars for which we have surface thermal (blackbody) luminosity observations, with the data collected in~\citet{Vigano2013} and updated in~\citet{Potekhin2018}. The simulations with Landau quantization effects show a small increase in the surface temperature, largely occuring at times after a few kyr. This effect is most noticeable by comparing the $B1$ configurations, which have a weaker initial field and thus lower overall Joule heating, so they reach equilibrium between Joule heating and neutrino plus surface photon emission at a lower crust temperature. This suggests that the thermal suppression of the dHvA oscillations of the magnetization and its derivatives is the key factor in why Landau quantization only marginally enhances the surface luminosity, especially for stronger initial magnetic fields: the Ohmic dissipation of these fields without Landau quantization is sufficient to maintain the crust at a high enough temperature such that the dHvA oscillations of magnetization have only a small amplitude and thus a small effect on magnetic field evolution and field dissipation. Based on our simulations, it seems clear that the small enhancement of the Joule heating provided by Landau quantization is unable to explain the observed surface luminosity of the hottest magnetars.

\begin{figure}
\centering
\includegraphics[width=0.98\linewidth]{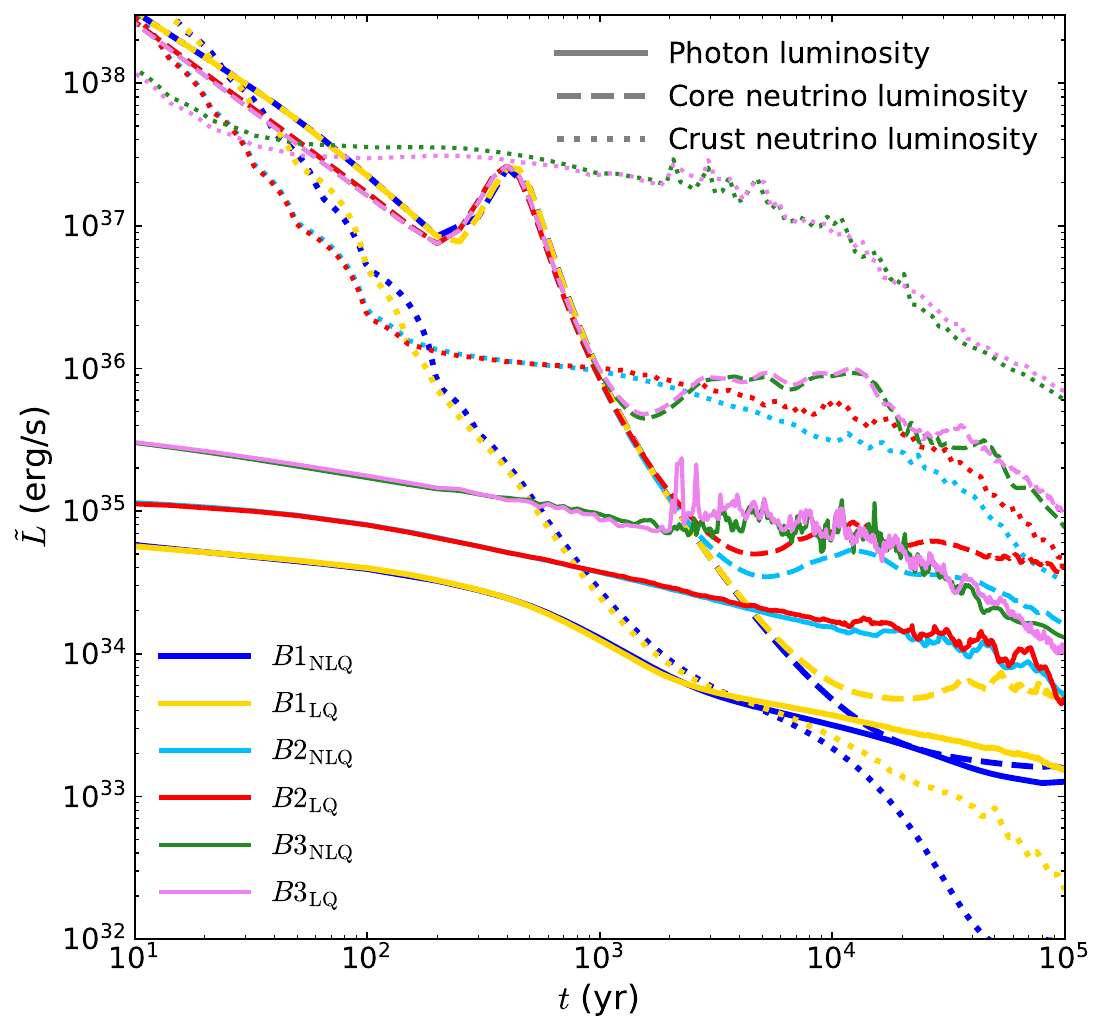}
\caption{Redshifted surface photon (solid line), core neutrino (dashed line) and crust neutrino (dotted line) luminosities as a function of time for four simulations defined in Table~\ref{tab:Simulations}. The core and crust neutrino emissivities are volume-integrated over the core and crust respectively.}
\label{fig:AllLuminosities}
\end{figure}

As the heat generated by Ohmic dissipation of the field is deposited throughout the crust, not all of it will reach the surface to be emitted as photons. We hence also examine the volume-integrated neutrino emissivities of the core and crust to determine how these are affected by the Landau quantization-enhanced Joule heating. This is shown in Figure~\ref{fig:AllLuminosities}, alongside the surface photon luminosity. The ``bump'' in the core neutrino luminosity at $t\approx 400$ yr is a result of the Cooper pair breaking and formation mechanism in the core, which is active when the core temperature is close to the critical temperature for $^3$P$_2$-paired neutrons. Landau quantization clearly increases the neutrino luminosity of both the crust and core by increasing the Joule heating in the crust and maintaining both crust and core at higher temperatures than otherwise. However, most of this heat goes into producing neutrinos instead of reaching the surface, so the photon luminosity is only marginally increased by Landau quantization. Once again, the enhancement of the neutrino luminosity due to Landau quantization-enhanced Joule heating is most prominent in the weakest initial field simulation $B1_{\rm LQ}$, with the core neutrino luminosity increased by a factor $\sim2$ and the crust neutrino luminosity by an order of magnitude for $t\gtrsim 50$ kyr. The increase in neutrino luminosity due to Landau quantization-enhanced heating for the simulations with stronger initial fields is more modest, with typical enhancements of both crust and core neutrino luminosities of order tens of percent after $t\sim 10$ kyr.

\subsection{Effect of crust impurities}

\begin{figure*}
\centering
\includegraphics[width=0.97\textwidth]{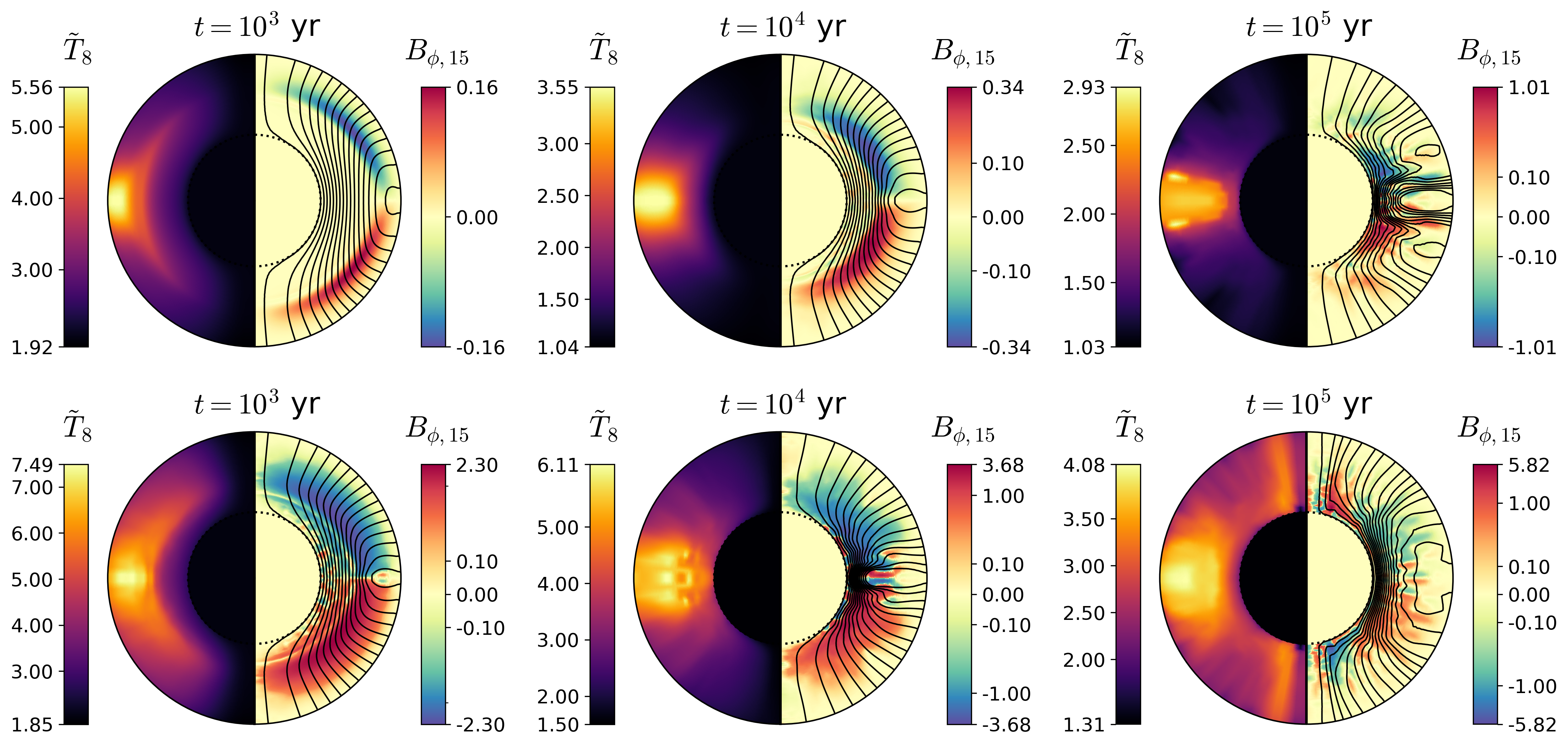}
\caption{Same as Figure~\ref{fig:MagnetothermalSimulationsLQ} except for the simulations with Landau quantization $B1_{\rm LQ,NI}$ (top row) and $B2_{\rm LQ,NI}$ (bottom row).}
\label{fig:MagnetothermalSimulationsNoImp}
\end{figure*}

The principal effect of the crust impurities is to enhance the Ohmic dissipation in the inner crust by decreasing the electrical conductivity; otherwise, the electrical conductivity in the inner crust is one-to-two orders of magnitude larger than in the outer crust at the same temperature. This suggests that, by reducing the Ohmic dissipation of the field and hence allowing the crust to reach lower temperatures, the effect of Landau quantization on neutron star magnetothermal evolution is significantly enhanced in the absence of crustal impurities/if the impurity parameter is much smaller than suggested in~\citet{Carreau2020a} ($Q\sim 10$'s in the inner crust).  Figure~\ref{fig:MagnetothermalSimulationsNoImp} shows snapshots of the magnetothermal evolution of simulations $B1_{\rm LQ,NI}$ and $B2_{\rm LQ,NI}$, identical to $B1_{\rm LQ}$ and $B2_{\rm LQ}$ but in the absence of crustal impurities and hence with lower electrical conductivity in the inner crust. As the effect of Landau quantization on configuration $B3$ was smaller than for $B1$ and $B2$, we do not consider it here. 

By comparing Figure~\ref{fig:MagnetothermalSimulationsNoImp} to Figure~\ref{fig:MagnetothermalSimulationsLQ}, we see that at $t=10^3$ yr, when the crust is still too hot for Landau quantization to greatly modify the magnetothermal evolution, the internal temperature and magnetic field are essentially the same with and without crustal impurities. The lack of impurities begins to affect the simulations more at $t=10^4$ yr, with $B1_{\rm LQ,NI}$ being slightly cooler than $B1_{\rm LQ}$ due to the reduced Joule heating, whereas $B2_{\rm LQ,NI}$ has a slightly warmer hot spot than $B2_{\rm LQ}$ and has a stronger small-scale toroidal field generated by the dHvA oscillations. This occurs because the inner crust is cooler in $B2_{\rm LQ,NI}$ due to the absence of impurities, and so greater amplitude dHvA oscillations can develop here and subsequently generate stronger small-scale field features and enhanced dissipation of said features. By $10^5$ years, we see that the lack of impurities has substantially changed the evolution, with the hot spot in $B1_{\rm LQ,NI}$ being $\sim 50$\% hotter than $B1_{\rm LQ}$ and the magnetic field showing much greater small-scale structure as a result of the dHvA oscillations. Similarly, the dHvA oscillations have greatly increased the amount of small-scale field structures in $B2_{\rm LQ,NI}$ compared to $B2_{\rm LQ}$, although the equatorial hot spot is at a lower temperature than in $B2_{\rm LQ}$. However, the crust on average is hotter in $B2_{\rm LQ,NI}$, implying a higher surface luminosity.

\begin{figure}
\centering
\includegraphics[width=0.98\linewidth]{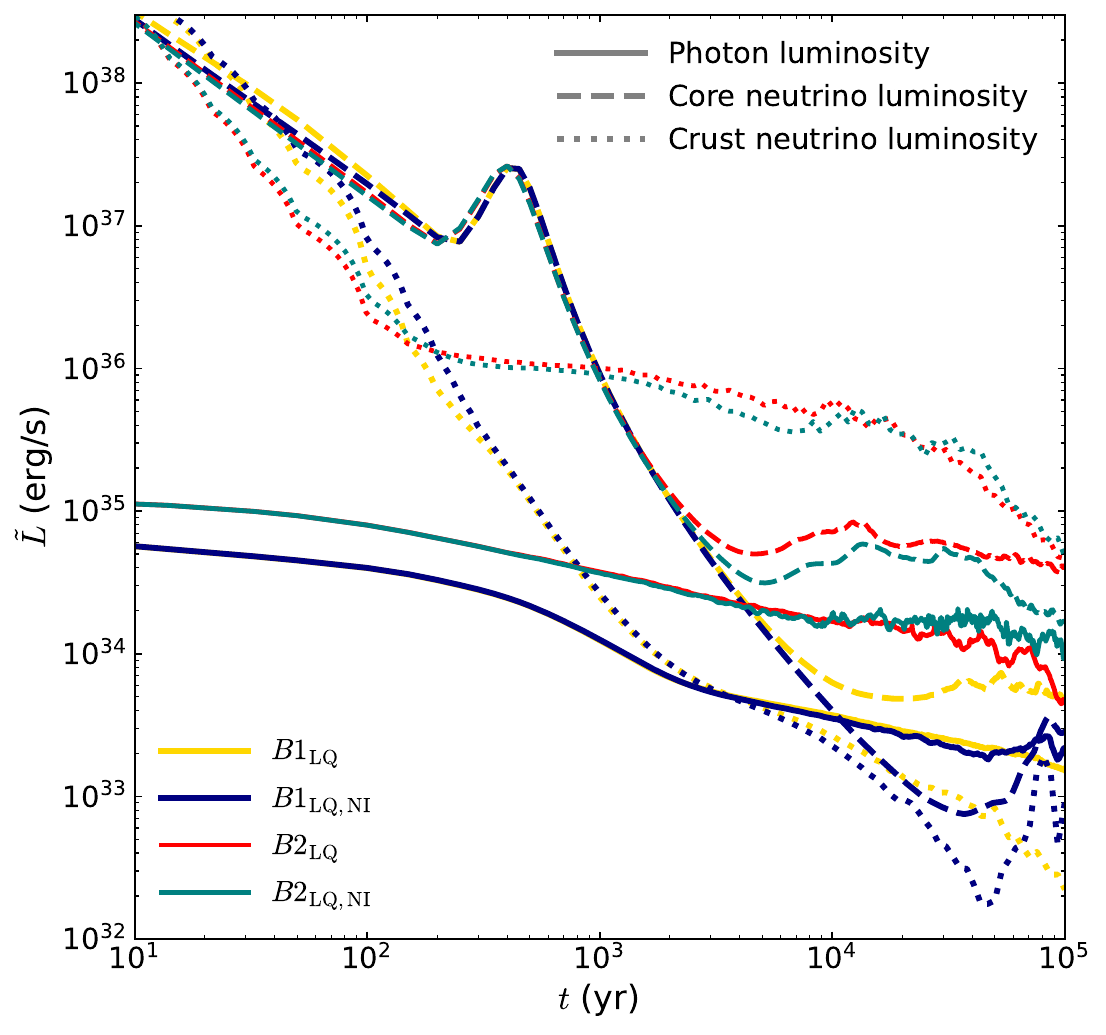}
\caption{Same as Figure~\ref{fig:AllLuminosities} except showing simulations $B1_{\rm LQ}$, $B1_{\rm LQ,NI}$, $B2_{\rm LQ}$ and $B2_{\rm LQ,NI}$.}
\label{fig:AllLuminositiesNoImp}
\end{figure}

That the late-time surface luminosity is stronger when Landau quantization is enabled and in the absence of impurities can be clearly seen by examining the luminosities as a function of time for simulations $B1_{\rm LQ,NI}$ and $B2_{\rm LQ,NI}$. The neutrino and surface photon luminosities of these simulations are shown in Figure~\ref{fig:AllLuminositiesNoImp}, alongside those of $B1_{\rm LQ}$ and $B2_{\rm LQ}$ for comparison. The results with and without impurities are broadly similar until $t\approx 3$ kyr, when the crust has cooled sufficiently for the dHvA oscillations to have significant amplitude. In simulation $B1_{\rm LQ,NI}$, the faster cooling crust results in a sharper decrease in the crust neutrino luminosity, before the resulting enhancement to the Joule heating from the larger amplitude dHvA oscillations at lower temperatures results in the crust heating up after $\sim 50$ kyr, with a greater surface photon and crust neutrino luminosity at the end of the simulation than $B1_{\rm LQ}$. However, the core thermal luminosity remains lower, since the reduced Joule heating in the inner crust due to the lack of impurities means that less heat from the crust flows into the core in $B1_{\rm LQ,NI}$. A similar comparison can be made for $B2_{\rm LQ,NI}$ and $B2_{\rm LQ}$ at $t\gtrsim 20$ kyr, with an enhancement of the surface photon luminosity in the former compared to the latter, a marginally increased crust neutrino luminosity, and a diminished core neutrino luminosity.

\section{Discussion and Conclusion}
\label{sec:Conclusion}

The magnetic fields in magnetar magnetospheres and crusts are sufficiently strong such that electron quantization into a single or relatively small number of Landau levels will occur. Landau quantization has been shown in previous studies to have dramatic effects on electron transport, neutrino emissivity and thermodynamic properties. In this study, for the first time we include the effects of Landau quantization in all of these properties in numerical simulations of magnetothermal evolution in a strongly magnetized neutron star crust using the code \texttt{QMFM}. Using a suite of simulations with different magnetic strengths, we have examined how these novel effects change the magnetothermal evolution, aiming to determine if they could sufficiently enhance the Joule heating of neutron star crusts to explain the high surface luminosity of the observed magnetars.

We find that the Landau quantization-induced enhancement of the Ohmic dissipation of the strongest neutron star magnetic fields is only modest, and the resulting Joule heating of the crust alone is unable to explain the observed surface luminosities of  the hottest magnetars. The enhancement to field dissipation provided by Landau quantization is greatest in the case when the Joule heating of the crust is weak, either because the initial field strength is lower or when crustal impurities are suppressed. The Landau quantization-enhanced Joule heating of the star can substantially increase the neutrino emissivity, although not nearly to the extent that this could be observed. The main reason why the Landau quantization-induced enhancement of the Ohmic dissipation is not more important is the thermal suppression of the dHvA oscillations: for fields with strengths $B\gtrsim 5\times10^{14}$ G, likely required to heat the star sufficiently to explain the hottest magnetars, the heating is too strong to allow the dHvA oscillations to grow to significant amplitude. It is when the Joule heating is the lowest--for weak initial fields $B\sim 10^{14}$ G-- that Landau quantization provides the greatest enhancement to the heating, though the resulting surface luminosities are still two orders of magnitude lower than the hottest-observed magnetars and an order of magnitude lower than those that we simulate for $B\sim 5\times10^{14}$ G fields without Landau quantization included.

An obvious topic of future study is to expand the variety of magnetic field configurations. If the initial field is stronger or includes higher multipole moments in the outer crust, the heat produced by its dissipation could more readily reach the surface, and the Landau quantization-induced enhancement of this heating could be of greater importance. We have also ignored core-threaded fields, which are more realistic as discussed in Section~\ref{sec:SimEquations}. These result in cooler stars~\citep{Vigano2013}, since heat from the core can readily flow to the surface along the magnetic field. Like the abscence of impurities in the crust did in our simulations, this could perhaps also increase the importance of Landau quantization-enhanced Joule heating, although whether the enhanced heating is sufficient to overcome the enhanced cooling requires additional study.

As mentioned in the introduction, we have also ignored the force balance on the crust and the possibility of plastic failure. We have considered very strong magnetic fields, especially in configurations $B2$ and $B3$, which may exert powerful stresses sufficient to cause the crust to undergo  failure~\citep{Li2016,Kojima2022,Gourgouliatos2022,Kojima2024}, thus violating the criterion for electron MHD. Plastic flow does not appear to suppress the Hall effect and has a smaller effect on poloidal field evolution~\citep{Gourgouliatos2021}, so neglecting it in our simulations of fields that have large initial poloidal components seems reasonable. However, most studies of plastic failure of the crust have assumed it occurs when magnetic and elastic stresses are balanced, which as~\citet{Bransgrove2025} argued is incorrect: for large-scale magnetic fields, the magnetic stress can be much larger than the elastic stress while the correct force balance condition, given by the divergences of these stresses, is satisfied. Hence the Landau quantization effects we study in this paper, through generating small-lengthscale but yet strong fields, may be well-placed to increase the divergence of the magnetic stress (i.e., the Lorentz force) sufficiently such that the required elastic stress to balance this exceeds the critical value for plastic failure. Plastic flow is an additional heat source which could also be important to the magnetothermal evolution of magnetars~\citep{Beloborodov2014,Li2016}; progress in determining the microphysics of this plastic flow, including the plastic flow viscosity which sets the heating rate, will enable its implementation in evolution codes such as \texttt{QMFM}. 

The seeming inability of Joule heating provided by strong fields alone, even including Landau quantization effects, alongside the disfavouring of a light element envelope in magnetars and the enhanced cooling provided by a physically realistic core-threaded field, suggests that a magnetospheric solution to the magnetar heating problem looks most promising at this stage.
\\
\\
PBR was supported by the Simons Foundation through a SCEECS postdoctoral fellowship (grant No. PG013106-02). The author would like to acknowledge Y. Levin and I. Wasserman for inspiration and insightful comments on this paper, and A. M, Beloborodov, A. J. Bransgrove and J. A. Pons for other useful discussions. Simulations were performed using the Perlmutter supercomputer, owned by the National Energy Research Scientific Computing Center (NERSC), a Department of Energy User Facility using NERSC award NP-ERCAP 0036106. All plots were made using the Python package \texttt{matplotlib}~\citep{Matplotlib}. The \texttt{QMFM} code and the scripts used to generate the plots in this paper are available at https://github.com/PRauNS/QMFM.

\vspace{5mm}

%\software{
%          \texttt{Matplotib} \citep{Matplotlib}, 
%          \texttt{NumPy} \citep{NumPy}
%          }

\appendix

\section{Temperature-dependent expressions for required thermodynamic functions}
\label{app:Thermodynamics}

In our simulations we compute the temperature and magnetic field-dependent thermodynamic functions $M$, $\chi_{\mu}$, $\mathcal{M}_{\mu}$, $\mathcal{M}_T$ and $c_{V,\rme}$. In the nonmagnetized case, the Sommerfeld expansion is sufficient to compute the finite temperature corrections to the $T=0$ Fermi gas thermodynamic functions. This approximation is only partially applicable to the $B\neq0$ case as discussed in~\citet{Rau2023}, so to avoid computationally expensive Fermi--Dirac integrals, we developed a set of different approximations to compute the required thermodynamic functions at finite temperature. The approximations for $M$, $\chi_{\mu}$ and $M_{\mu}$ and their derivations are described in detail in the Appendix of RW25, so he were merely state the approximate forms for $\mathcal{M}_T$ and $c_{V,\rme}$, which are derived in identical fashion. We work in Gaussian units and set $\hbar=c=k_{\rm{B}}=1$.

\subsection{Low-temperature approximations}

All of the $B$-dependent thermodynamic functions of interest to us can be calculated starting from the grand potential density $\Omega_{\rme}(B,\mu_e,T)$ for the magnetized electrons. $\mathcal{M}_T$ and $c_{V,\rme}$, the functions required in this work that were not used in RW25, are both second-order partial derivatives of $\Omega_{\rme}$, or first-order partial derivatives of the electron entropy density $s_{\rme}$. For completeness, we include the expressions for $s_{\rme}$, from which $\mathcal{M}_T$ and $c_{V,\rme}$ are calculated. 

At $T=0$, electrons will occupy Landau levels up to $n_{\text{max}}=\lfloor p_F^2/(2eB)\rfloor$, where $p_F=\sqrt{\mu_{\rme}^2-m_{\rme}^2}$. At low but nonzero temperatures $T\leq eB/(2\pi^2\mu_{\rm e})$, only a few Landau levels above $n=n_{\text{max}}$ will be occupied, and the Landau levels below $n_{\text{max}}$ are nearly unaffected by the temperature. To calculate $s_{\rme}$, $\mathcal{M}_T$ and $c_{V,\rme}$ for low temperature, we thus use the Sommerfeld expansion around $T=0$ for $\Omega_{\rme}$ and take partial derivatives for $n<n_{\text{max}}$, and then include distinct corrections only for $n=n_{\text{max}}$ and $n_{\text{max}}+1$, where the Sommerfeld expansion breaks down. Defining $E_{Fn}=\sqrt{\mue^2-m_{\rm e}^2-2eBn}$, $m_n=\sqrt{m_{\rme}^2+2eBn}$ and $n'=n_{\rm max}-1$, for $n=0,1,...n'$ we use the Euler--Maclaurin summation formula and the Sommerfeld expansion to write
\begin{align}
%\left(s_{\rme}\right)_{n<n_{\rm max}}=-\left(\frac{\partial\Omega_{\rme}}{\partial T}\right)_{n<n_{\rm max}}={}&\frac{eB\mue T}{3}\Bigg[ \frac{1}{2p_F} - \frac{E_{Fn'}-E_{F1}}{eB} + \frac{1}{2E_{Fn'}} + \frac{1}{2E_{F1}} + \mathcal{B}_2\frac{eB}{2}\left(\frac{1}{E_{Fn'}^3}-\frac{1}{E_{F1}^3}\right) 
\left(s_{\rme}\right)_{n<n_{\rm max}}=-\left(\frac{\partial\Omega_{\rme}}{\partial T}\right)_{n<n_{\rm max}}={}&\frac{eB\mue T}{6}\Bigg[ \frac{1}{p_F} - \frac{2\left(E_{Fn'}-E_{F1}\right)}{eB} + \frac{1}{E_{Fn'}} + \frac{1}{E_{F1}} + \mathcal{B}_2eB\left(\frac{1}{E_{Fn'}^3}-\frac{1}{E_{F1}^3}\right) 
\nonumber
\\
{}& \qquad\qquad+\mathcal{B}_4\frac{5(eB)^3}{4}\left(\frac{1}{E_{Fn'}^7}-\frac{1}{E_{F1}^7}\right)\Bigg],
\\
\left(\mathcal{M}_T\right)_{n<n_{\rm max}}=\left(\frac{\partial s_{\rme}}{\partial B}\right)_{n<n_{\rm max}}={}&
\frac{T\mue}{3}\Bigg[\frac{1}{2p_F}+\frac{n'}{E_{Fn'}}-\frac{1}{E_{F1}}
+\frac{1}{2}\left(\frac{E_{Fn'}^2 + eBn'}{E_{Fn'}^3}+\frac{E_{F1}^2 + eB}{E_{F1}^3}\right)
\nonumber
\\
{}&+\mathcal{B}_2\frac{eB}{2}\left(\frac{2E_{Fn'}^2 + 3eBn'}{E_{Fn'}^5}-\frac{2E_{F1}^2 + 3eB}{E_{F1}^5}\right)
+\mathcal{B}_4\frac{5(eB)^3}{8}\left(\frac{4E_{Fn'} + 7eBn'}{E_{Fn'}^9}-\frac{4E_{F1} + 7eB}{E_{F1}^9}\right)\Bigg],
\\
\left(c_{V,\rme}\right)_{n<n_{\rm max}}=\left(\frac{\partial s_{\rme}}{\partial T}\right)_{n<n_{\rm max}}={}&\frac{\left(s_{\rme}\right)_{n<n_{\rm max}}}{T}
+ \frac{7\pi^2eB\mu_eT^2}{10}\Bigg[\frac{m_{\rme}^2}{2p_F^5} - \frac{1}{3eB}\left(\frac{2E_{Fn'}^2 - m_{n'}^2}{E_{Fn'}^3}-\frac{2E_{F1}^2 - m_1^2}{E_{F1}^3}\right) + \frac{1}{2}\left(\frac{m_{n'}^2}{E_{Fn'}^{5}}+\frac{m_1^2}{E_{F1}^5}\right)
\nonumber
\\
{}& + \mathcal{B}_2\frac{eB}{2}\left(\frac{2E_{Fn'}^2 + 5m_{n'}^2}{E_{Fn'}^7}-\frac{2E_{F1}^2 + 5m_1^2}{E_{F1}^7}\right)
+ \mathcal{B}_4\frac{35(eB)^3}{8}\left(\frac{2E_{Fn'}^2 + 3m_{n'}^2}{E_{Fn'}^{11}}-\frac{2E_{F1}^2 + 3m_1^2}{E_{F1}^{11}}\right)\Bigg].
\end{align}
$\mathcal{B}_2=1/6$ and $\mathcal{B}_4=-1/30$ are Bernoulli numbers. For $n=n_{\text{max}}$ and $n_{\text{max}}+1$, we use the following approximations:
\begin{align}
\left(s_{\rme}\right)_{n\geq n_{\rm max}}={}&\frac{eB}{2\pi^2T}\sum_{n=n_{\rm max}}^{n_{{\rm max}+1}}g_nm_n^2\left[2\mathcal{G}_1(a,b)-\mathcal{G}_2(a,b)-\frac{b}{a}\mathcal{I}_1(a,b)\right],
\\
\left(\mathcal{M}_T\right)_{n\geq n_{\rm max}}={}&\frac{e}{2\pi^2T}\sum_{n=n_{\rm max}}^{n_{{\rm max}+1}}g_n\left[m_n^2\left(2\mathcal{G}_1(a,b)-\mathcal{G}_2(a,b)-\frac{b}{a}\mathcal{I}_1(a,b)\right)+eBn\left(\mathcal{H}_2(a,b)-\frac{b}{a}\mathcal{I}_2(a,b)\right)\right],
\\
\left(c_{V,\rme}\right)_{n\geq n_{\rm max}}={}&\frac{eB}{2\pi^2T^2}\sum_{n=n_{\rm max}}^{n_{{\rm max}+1}}g_nm_n^2\left[2\mathcal{G}_1(a,b)-\mathcal{G}_2(a,b)-\left(\left(\frac{b}{a}\right)^2+1\right)\mathcal{H}_2(a,b)+\frac{2b}{a}\left(\mathcal{I}_2(a,b)-\mathcal{I}_1(a,b)\right)\right].
\end{align}
where $g_n=(2-\delta_{n,0})$, $a=m_n/T$ and $b=\mu_{\rme}/T$. The functions $\mathcal{G}_1$, $\mathcal{G}_2$, $\mathcal{H}_2$, $\mathcal{I}_1$, and $\mathcal{I}_2$ and their approximations for different parts of the $(a,b)$ phase space are provided in the Appendix of RW25. In the $b\gg a$ limit, we calculate $s_{\rme}$, $\mathcal{M}_T$ and $c_{V,\rme}$ analogously to Eq.~(A13),~(A20) and~(A21) of RW25; that is, using the $T=0$ limit plus the Sommerfeld expansion.

\subsection{High-temperature approximations}
\label{app:HighTApprox}

At high temperatures $T> eB/(2\pi^2\mu_{\rm e})$, we use an approximation based on splitting of $\Omega_e$ into nonmagnetic $\Omega_{\rm e,0}$, regular $\Omega_{\rm e,\text{reg}}$ and oscillatory $\Omega_{\rm e,\text{osc}}$ parts using the Poisson summation formula as described in~\citet{Elmfors1993}. Nonzero temperature is accounted for by including the nonzero temperature correction factors in $\Omega_{\rm e,\text{osc}}$, through Sommerfeld expansion terms for $\Omega_{\rm e,0}$ where they are nonzero, and Sommerfeld expansion terms for partial derivatives of $\Omega_{\rm e,\text{reg}}$ with respect to $T$. The resulting approximations for $s_e$, $\mathcal{M}_T$ and $c_{V,\rme}$ are
\begin{subequations}
\begin{align}
s_{\rm e} = {}& \frac{\mue p_F T}{3} + \frac{eB^{3/2}}{2\pi^2}\sum_{p=1}^{\infty}\frac{R_{2,T,p}}{p^{3/2}}\cos\left(\frac{\pi}{4}-\pi p\frac{p_F^2}{eB}\right) -\frac{\sqrt{eB}\mue T}{6\sqrt{\pi}}\mathcal{K}_1\left(\frac{p_F^2}{eB}\right),
\label{eq:HighTs}
\\
\mathcal{M}_T = {}& -\frac{e}{2\pi\sqrt{eB}}\sum_{p=1}^{\infty}\frac{R_{2,T,p}}{\sqrt{p}}\left[\mue^2\sin\left(\frac{\pi}{4}-\pi p\frac{p_F^2}{eB}\right)-\frac{3eB}{2\pi p}\cos\left(\frac{\pi}{4}-\pi p\frac{p_F^2}{eB}\right)\right] + \frac{e\mue T}{12\sqrt{\pi eB}}\left[\mathcal{K}_1\left(\frac{p_F^2}{eB}\right) - \frac{2p_F^2}{eB}\mathcal{K}_2\left(\frac{p_F^2}{eB}\right)\right],
\\
c_{V,\rme} = {}& \frac{\mue p_F}{3} - \frac{(eB)^{3/2}}{2\pi^2T}\sum_{p=1}^{\infty}\frac{R_{3,T,p}}{p^{3/2}}\cos\left(\frac{\pi}{4}-\pi p\frac{p_F^2}{eB}\right) - \frac{\sqrt{eB}\mue}{6\sqrt{\pi}}\mathcal{K}_1\left(\frac{p_F^2}{eB}\right),
\label{eq:HighTc_V}
\end{align}
\end{subequations}
where we retain only the leading order or a few subleading order terms in the ratio $p_F^2/eB\gg 1$. The auxiliary function $\mathcal{K}_p$ is the same as RW25: 
\begin{align}
\mathcal{K}_p(x) ={}& \int_0^{\infty}\frac{{\rm d}y}{y^{5/2-p}}\left[y\text{coth}y-1\right]{\rm e}^{-yx}.
\end{align}
We cut off the sums over $p$ in Eq.~(\ref{eq:HighTs}--\ref{eq:HighTc_V}) at $p=5$ and obtain very accurate results in the parameter range of interest.

The functions $R_{2,T,p}$ and $R_{3,T,p}$ are analogs to $R_{T,p}$ for functions involving one and two partial derivatives with respect to $T$, and are obtained following~\citet{Shoenberg1984}, Chpt. 2.3.7. We show the calculation of $R_{2,T,p}$ in detail. The oscillatory contribution to the entropy density $s_{\rm e}$ in the nonzero temperature case is defined as
\begin{equation}
s_{\rm e,\text{osc}} 
=-\frac{(eB)^{3/2}}{2\pi^3}\sum_{p=1}^{\infty}\frac{1}{p^{3/2}}\int_{m_{\rm e}}^{\infty}{\rm d}E\frac{\partial f_{F}}{\partial T}\sin\left(\frac{\pi}{4}-\frac{\pi p}{eB}\left(E^2-m_{\rm e}^2\right)\right)
=\frac{(eB)^{3/2}}{2\pi^3}\sum_{p=1}^{\infty}\frac{1}{p^{3/2}}\int_{m_{\rm e}}^{\infty}{\rm d}E\frac{E-\mue}{T}\frac{\partial f_{F}}{\partial E}\sin\left(\psi(E)\right).
\end{equation}
The effect of the thermal phase smearing is to replace the integral over $E$ with the following ratio
\begin{equation}
I_2 = \left[\int_{-\infty}^{\infty}\rmd\phi\sin(\psi(E)+\phi)\frac{\phi}{\lambda}\mathcal{D}\left(\frac{\phi}{\lambda}\right)\right]\Bigg/ \int_{-\infty}^{\infty}{\rm d}\phi \mathcal{D}\left(\frac{\phi}{\lambda}\right),
\end{equation}
where $\mathcal{D}(\phi/\lambda)=\partial f_{F}/\partial E$ for $f_F(E,\mue,T)=(\exp([E-\mue]/T)+1)^{-1}$, $\lambda = 2\pi^2 p\mue T/(eB)$ and $\phi/\lambda = (E-\mue)/T$. For $\psi(E)$ approximately independent of $\phi$, we can write
\begin{equation}
I_2 = \text{Re}\left[-i\rme^{i\psi(E)}\int_{-\infty}^{\infty}{\rm d}\phi\rm e^{i\phi}\frac{\phi}{\lambda}\mathcal{D}\left(\frac{\phi}{\lambda}\right)\right]\Bigg/ \int_{-\infty}^{\infty}{\rm d}\phi \mathcal{D}\left(\frac{\phi}{\lambda}\right)=\pi\text{csch}(\lambda\pi)(\pi\lambda\text{coth}(\pi\lambda)-1)\cos(\psi(E)),
\end{equation}
from which we extract 
\begin{equation}
R_{2,T,p}\equiv \text{csch}(\lambda\pi)\left[\pi\lambda\text{coth}(\pi\lambda)-1\right].
\end{equation}
Partial differentiation of $s_{\rm e}$ with respect to $B$ gives $\mathcal{M}_T$, which hence also includes the $R_{2,T,p}$ factor. An analogous calculation for $c_{V,\rme}(B)=\partial^2\Omega_{\rm e}/\partial T^2$ gives
\begin{equation}
R_{3,T,p}\equiv \frac{\pi\lambda}{2}\text{csch}^3(\lambda\pi)\left[3\pi\lambda+\pi\lambda\text{cosh}(\pi\lambda)-2\text{sinh}(2\pi\lambda)\right].
\end{equation}

The approximations for $\mathcal{M}_T$ and $c_{V,\rme}$ were compared to their exact forms, computed by calculating the full Fermi--Dirac integrals. Over a parameter space range $10^{14}\text{ G}<B<5\times10^{16}$ G, $10\text{ MeV}<\mue<80$ MeV and $5\times10^{7}\text{ K}<T<5\times10^9$ K, the mean relative error between the approximations and exact values is $\lesssim $1\%. The maximum relative error between approximate and exact values is 12\% for $\mathcal{M}_T$ and 14\% for $c_{V,\rme}$.

\bibliographystyle{aasjournal}
\bibliography{library,textbooks,librarySpecial}

\end{document}